\documentclass[journal,twoside]{IEEEtran}
\usepackage{graphicx,amsmath,amssymb,color}
\usepackage{verbatim}
\usepackage[noadjust]{cite}
\usepackage{array}
\usepackage{cases}
\usepackage{url}
\usepackage{multirow}
\newcommand{\ket}[1]{\left\vert #1 \right\rangle}
\newcommand{\bra}[1]{\left\langle #1 \right\vert}

\newtheorem{theorem}{Theorem}[section]
\newtheorem{proposition}[theorem]{Proposition}
\newtheorem{corollary}[theorem]{Corollary}
\newtheorem{lemma}[theorem]{Lemma}

\begin{document}
\title{High-Rate Quantum Low-Density Parity-Check Codes Assisted by Reliable Qubits}
\author{Yuichiro~Fujiwara,~\IEEEmembership{Member,~IEEE}, Alexander Gruner, and Peter Vandendriessche,~\IEEEmembership{Member,~IEEE}%
\thanks{This work was supported by JSPS (Y.F.), LGFG Baden-W\"{u}rttemberg (A.G.), and FWO (P.V.).
The third author is support by a PhD fellowship of the Research Foundation - Flanders (FWO).}%
\thanks{Y. Fujiwara is with the Division of Physics, Mathematics and Astronomy, California Institute of Technology, MC 253-37, Pasadena, CA 91125 USA
{(email: yuichiro.fujiwara@caltech.edu)}.}%
\thanks{A. Gruner is with the Mercedes-Benz Bank AG, Siemensstra{\ss}e 7, 70469 Stuttgart, Germany
{(email: alexander.gruner@uni-tuebingen.de)}.}%
\thanks{P. Vandendriessche is with the Department of Mathematics, Ghent University, Krijgslaan 281 - S22, 9000 Ghent, Belgium
{(email: pv@cage.ugent.be)}.}
}
\markboth{IEEE transactions on Information Theory,~Vol.~x, No.~xx,~month~year}
{Fujiwara, Gruner, and Vandendriessche: High-rate quantum low-density parity-check codes assisted by reliable qubits}

%\pubid{0000--0000/00\$00.00~\copyright~2007 IEEE}

\maketitle

\begin{abstract}
Quantum error correction is an important building block for reliable quantum information processing.
A challenging hurdle in the theory of quantum error correction is that
it is significantly more difficult to design error-correcting codes with desirable properties for quantum information processing
than for traditional digital communications and computation.
A typical obstacle to constructing a variety of strong quantum error-correcting codes is the complicated restrictions imposed on the structure of a code.
Recently, promising solutions to this problem have been proposed in quantum information science, where in principle any binary linear code
can be turned into a quantum error-correcting code by assuming a small number of reliable quantum bits.
This paper studies how best to take advantage of these latest ideas to construct desirable quantum error-correcting codes of very high information rate.
Our methods exploit structured high-rate low-density parity-check codes available in the classical domain
and provide quantum analogues that inherit their characteristic low decoding complexity and high error correction performance even at moderate code lengths.
Our approach to designing high-rate quantum error-correcting codes
also allows for making direct use of other major syndrome decoding methods for linear codes,
making it possible to deal with a situation where promising quantum analogues of low-density parity-check codes are difficult to find.
\end{abstract}


\IEEEpeerreviewmaketitle

\section{Introduction}
\IEEEPARstart{Q}{uantum} error-correcting codes are schemes that recover the original quantum information
when the quantum states of quantum bits, or \textit{qubits}, carrying the information
are transformed by unintended quantum operations, namely \textit{quantum noise} \cite{Nielsen:2000}.
As is the case with traditional information processing, it is vital to suppress the effect of quantum noise when processing quantum information.
The role of error correction is particularly crucial in the quantum domain because qubits are expected to be highly vulnerable to environmental noise
in practical and realistic situations.

While the importance of reliability is apparent,
there had been doubts about the existence of a viable scheme for error correction in the quantum domain
until the discovery of the famous $9$-qubit code \cite{Shor:1995} and $7$-qubit code \cite{Steane:1996} in the mid 1990's.
These findings ignited intensive and rapidly progressing research on error correction for quantum information.
In fact, various types of quantum error-correcting code are now known including the celebrated stabilizer codes \cite{Gottesman:1997,Calderbank:1998},
which constitute a very general class encompassing the first two quantum error-correcting codes,
and codeword stabilized codes \cite{Cross:2009}.
Small quantum error-correcting codes, such as the perfect $5$-qubit quantum error-correcting code \cite{Laflamme:1996},
have been experimentally realized as well
\cite{Cory-et-al.:1998,Knill:2001a,Chiaverini-et-al.:2004,Boulant:2005b,Lu-et-al.:2008,Aoki-et-al.:2009,Schindler-et-al.:2011,
Moussa:2011,Reed-et-al.:2012,Yao-et-al.:2012a,Zhang:2012b,Zhang:2012,Bartz-et-al.:2014,Bell:2014}.

However, this remarkable progress does not mean that the theory of quantum error correction has become as mature as classical coding theory.
It would be more accurate to say that we just started finding ways to realize quantum error correction
while cleverly circumventing challenging obstacles imposed by quantum mechanical phenomena.

For instance, while the stabilizer formalism developed in \cite{Gottesman:2010}
has given rise to a wide range of quantum error-correcting codes,
one of the theoretically challenging problems with this approach is that the admissible structures of a code are severely restricted
when compared to the freedom we have in classical code design.
The fact that there are only few successful general frameworks for quantum code design also limits the variety of quantum error-correcting codes,
which is a crucial problem because actual realizations of large-scale quantum information processing is expected to demand various types of peculiar requirement.

One effective way to overcome the limitations and difficulties in the quantum domain is 
to develop a fresh and quantum mechanically valid framework
that makes it possible to directly import a wider range of classical coding theory to the quantum regime.
The \textit{entanglement-assisted stabilizer formalism} is a major breakthrough in this direction,
where one may fully exploit any binary or quaternary linear code over the binary field $\mathbb{F}_2$ or the finite field $\mathbb{F}_4$ of order four respectively
for correcting errors on qubits as long as there is an adequate supply of maximally entangled noiseless qubits to assist quantum error correction \cite{Brun:2006}.
A pair of maximally entangled qubits is called an \textit{ebit}.
Entanglement-assisted quantum error-correcting codes can be regarded as generalized stabilizer codes
in that those requiring no ebit are exactly the standard stabilizer codes;
if a linear code can not be turned into a quantum error-correcting code through the standard stabilizer formalism,
one may still exploit it by assuming that some amount of quantum resources can be shared through a noiseless channel as ebits
to help encode and decode noisy qubits.

A major drawback of entanglement assistance is that completely noiseless qubits are extremely difficult to provide in a practical quantum device.
This disadvantage is particularly pronounced in the context of storing quantum information, where
the information source and sink may not be spatially distant but are separate in the time domain.
This characteristic of entanglement-assisted quantum error-correcting codes led to a series of research trying to
identify excellent linear codes which can be imported by relying only on a tiny number of ebits
\cite{Hsieh:2007,Wilde:2008,Hsieh:2009,Dong:2009,Djordjevic:2010,Fujiwara:2010e,Wilde:2010,Wilde:2010d,Hsieh:2011,Lai:2012,Fujiwara:2013b,Guo:2013}.
Playing a crucial role in these theoretical results is the assumed future technology of manipulating a small number of qubits with extreme reliability
to realize perfect and stable ebits.

Very recently, a framework that significantly reduces the burden of providing extreme reliability has been proposed,
where any binary or quaternary linear codes over $\mathbb{F}_2$ or $\mathbb{F}_4$ respectively can be fully exploited
as long as we can provide auxiliary qubits that are only subject to a restricted quantum error model \cite{Fujiwara:2013c}.
This framework takes advantage of the fact that while realizing completely noiseless qubits is a very difficult task,
not every kind of quantum error is equally difficult to suppress through technical development on hardware.
For instance, it is known that one can correct any type of quantum error in the standard general error model
if two particular types of error, called a bit error and phase error, can be corrected under the assumption that both may happen on the same qubit \cite{Nielsen:2000}.
However, phase errors due to dephasing are expected to be far more likely than bit errors in many actual quantum devices \cite{Ioffe:2007},
which implies that bit errors would be far less problematic.
In the newer framework, one may choose an error model in such a way that most qubits can suffer from bit errors and phase errors
while only phase errors may occur on a small number of auxiliary qubits.
Hence, unlike the entanglement-assisted stabilizer formalism, which requires completely noiseless auxiliary qubits,
the newer framework only needs more easily achievable ``less noisy'' ones.

With all these advances in this field,
one may think that the problem of severely restricted structures of quantum error-correcting codes is largely solved.
The caveat is that the statement that a given linear code $\mathcal{C}$ can be imported as a stabilizer code, entanglement-assisted quantum error-correcting code
or one that is assisted by less noisy qubits only means that
$\mathcal{C}$ admits a suitable parity-check matrix that is exploitable for quantum error correction.
In other words, of all distinct parity-check matrices that define one same linear code $\mathcal{C}$, only some special ones are usable.

To illustrate this problem, consider linear codes that greatly benefit from parity-check matrices in particular form in the classical domain.
For instance, \textit{low-density parity-check} (LDPC) \textit{codes} are linear codes
that admit parity-check matrices with a small number of nonzero entries such that iterative decoding performs well \cite{MacKay:2003}.
They are among the state-of-the-art error-correcting codes in classical coding theory
in the sense that well-designed LDPC codes almost achieve the Shannon limit over some channels and have remarkably low decoding complexity.
Since we have a means to import the theory of linear codes into the quantum domain,
LDPC codes constitute very promising ingredients for quantum error correction.
However, whether a given linear code is qualified as an excellent LDPC code depends on
whether it has a parity-check matrix suitable for iterative decoding.
This implies that whether we may have a quantum counterpart that inherits the attractive characteristics of a given LDPC code
depends on whether its particular parity-check matrix suitable for iterative decoding
is compatible with the chosen method for turning a linear code into a quantum error-correcting code.

The purpose of the present paper is to give insight into how best to exploit the recently proposed frameworks
for quantum error correction assisted by reliable qubits when the restrictions on parity-check matrices must be taken into account.
In particular, we focus on the case when excellent LDPC codes are used
to achieve high performance in both decoding complexity and error correction.
To take full advantage of auxiliary qubits while keeping our work well-focused,
we aim to construct quantum error-correcting codes with a few other properties that would be desirable in various situations.

An $[[n,k]]$ quantum error-correcting code of \textit{length} $n$ and \textit{dimension} $k$ encodes $k$-qubit information into $n$ physical qubits,
where the two nonnegative integers $n$ and $k$ satisfy the condition that $n > k \geq 0$.
The first property we aim for is a very high information rate in the absolute sense,
which means that we would like an $[[n,k]]$ quantum error-correcting code with $k$ close to $n$.
In addition to this condition, we strive to restrict ourselves to quantum error-correcting codes of modest and realistic length.
Thus, we do not consider the case when the parameter $n$ is an unrealistically large integer or the purely theoretical case of $n$ approaching infinity.
The feasibility of implementing our quantum error-correcting codes is also of importance.
For this reason, we only allow a very small number of reliable auxiliary qubits.

To illuminate the potential of our approach,
we aim for simultaneously satisfying the demanding conditions described above while achieving high error correction performance comparable to
what would be attainable in a hypothetical situation where some of the best known classical LDPC codes were freely available for quantum error correction
without the limitation on the structure of parity-check matrices.
As we will see later, carefully designed quantum LDPC codes can achieve this goal through assisted quantum error correction.
Furthermore, a brief discussion at the end of this paper will show how assisted quantum error correction with less noisy qubits,
if exploited with a different decoding method for linear codes, may remain successful
in a situation where excellent quantum LDPC codes are difficult to construct.

In the next section we briefly review quantum error-correcting codes assisted by reliable qubits.
Section \ref{LDPC} discusses the use of LDPC codes of high rate for quantum error correction
through the recently proposed frameworks.
We examine the performance of our quantum LDPC codes through simulations in Section \ref{sim}.
Concluding remarks including a brief discussion on how to apply assisted quantum error correction to other decoding methods
are given in Section \ref{conc}.

\section{Quantum Error Correction with Reliable Auxiliary Qubits}\label{review}
In this section we give a brief review of how reliable auxiliary qubits help correct quantum errors.
For the basics of quantum information theory, we refer the reader to \cite{Nielsen:2000}.
All facts in classical coding theory we use in this section can be found in \cite{Huffman:2003b}.

As usual, by a binary linear $[n,k,d]$ code, we mean a $k$-dimensional subspace $\mathcal{C}$ of the $n$-dimensional vector space over $\mathbb{F}_2$
in which a nonzero vector with the smallest number of nonzero entries has exactly $d$ nonzero entries, that is,
$\min\{\operatorname{wt}(\boldsymbol{c}) \mid \boldsymbol{c} \in \mathcal{C}, \boldsymbol{c}\not=0\}=d$.
Because we only consider a binary code, we omit the term binary when referring to linear codes and LDPC codes.
As stated earlier, an $[[n,k]]$ quantum error-correcting code encodes $k$ logical qubits into $n$ physical qubits,
which is analogous to a linear $[n,k,d]$ code in the sense that the classical code encodes $k$ logical bits into $n$ physical bits.

An important fact in the quantum domain is that, through a process called \textit{discretization},
an error correction scheme can correct any general quantum error on one qubit
if it can correct the effects of the Pauli operators $X, Z$ and their product $XZ$,
where the operator $X$ corresponds to the \textit{bit error} on one qubit while $Z$ represents the \textit{phase error}  \cite{Nielsen:2000}.
Similarly, quantum errors on multiple qubits can be corrected if the corresponding transformation by a combination of $X$, $Z$ and both at the same time
on each of the affected qubits is detected and reversed.

The quantum error-correcting codes we consider in this paper also take advantage of discretization.
Hence, without loss of generality, we always assume that
a quantum channel may introduce on each qubit only a bit error, a phase error or both at the same time
as a quantum error during information transmission unless otherwise stated.

The rest of this section is divided into two subsections.
Section \ref{lessnoisy} presents the basics of the framework for quantum error correction given in \cite{Fujiwara:2013c}
that is assisted by qubits on which only one particular kind of quantum error may occur.
We briefly review in Section \ref{entanglement} the entanglement-assisted stabilizer formalism developed in \cite{Brun:2006}
which uses completely noiseless qubits.

\subsection{Less Noisy Auxiliary Qubits}\label{lessnoisy}
Here we give the basics of quantum error correction assisted by less noisy qubits from the viewpoint of classical coding theory.
The following is the tool we use to import linear codes.
\begin{theorem}[\cite{Fujiwara:2013c}]\label{binaryonlyZ}
If there exists a linear $[n,k,d]$ code,
then there exist unitary operations that encode $k$ logical qubits into $2n-k$ physical qubits and correct up to $\lfloor\frac{d-1}{2}\rfloor$ quantum errors
under the assumption that a fixed set of $2(n-k)$ physical qubits may experience phase errors but no bit errors.
\end{theorem}

Roughly speaking, the above theorem says that any linear $[n,k,d]$ code, which corrects errors on up to $\lfloor\frac{d-1}{2}\rfloor$ bits,
can be turned into a $[[2n-k,k]]$ quantum error-correcting code
that corrects quantum errors on up to $\lfloor\frac{d-1}{2}\rfloor$ qubits as long as predetermined $2(n-k)$ qubits are only subject to phase errors.
Note that if the original linear $[n,k,d]$ code is of sufficiently high rate, the $2(n-k)$ auxiliary qubits consist of only a small fraction of the $2n-k$ physical qubits.

A particularly useful fact regarding this type of quantum error correction is that
we can employ decoding methods for linear codes based on error syndromes.
We formulate this most fundamental part of our approach in the form of a theorem below.
\begin{theorem}\label{lessnoisylemma}
Let $\mathcal{C}$ be a linear $[n,k,d]$ code.
Assume that $2n-k$ physical qubits $q_i$, $0 \leq i \leq 2n-k-1$, are sent through a noisy quantum channel in which
the first $2(n-k)$ qubits $q_i$, $0 \leq i \leq 2(n-k)-1$ are only subject to phase errors while
the remaining $k$ qubits $q_i$, $2(n-k) \leq i \leq 2n-k-1$ are subject to both bit errors and phase errors.
Define a pair $\boldsymbol{e}_0 = (e_0,\dots,e_{n-1}), \boldsymbol{e}_1 = (e'_0,\dots,e'_{n-1}) \in \mathbb{F}_2^{n}$
of $n$-dimensional vectors
such that for $0 \leq i \leq n-k-1$,
$e_i = 1$ if a phase error occurred on $q_i$ and $e_i = 0$ otherwise,
such that for $n-k \leq i \leq n-1$,
$e_i = 1$ if a bit error occurred on $q_{i+n-k}$ and $e_i = 0$ otherwise,
and such that for $0 \leq i \leq n-1$, $e'_{i} = 1$ if a phase error occurred on $q_{i+n-k}$ and $e'_i = 0$ otherwise.
Let $H$ be a parity-check matrix of $\mathcal{C}$ in standard form.
There exists a $[[2n-k,k]]$ quantum error-correcting code
that allows for retrieving classical information about quantum errors
in the form of a pair $\boldsymbol{s}_0, \boldsymbol{s}_1 \in \mathbb{F}_2^{n-k}$ of $(n-k)$-dimensional vectors
such that $\boldsymbol{s}_0 = H\boldsymbol{e}_0^T$ and $\boldsymbol{s}_1 = H\boldsymbol{e}_1^T$.
\end{theorem}

Note that the binary vectors $\boldsymbol{e}_0$ and $\boldsymbol{e}_1$ in the above theorem specify
what type of quantum error occurred on which qubit.
The correspondence between each bit of the error vectors $\boldsymbol{e}_0$, $\boldsymbol{e}_1$
and the type and location of each quantum error is summarized in Fig.\ \ref{figcorrespondence}.
\setlength{\unitlength}{8.4mm}
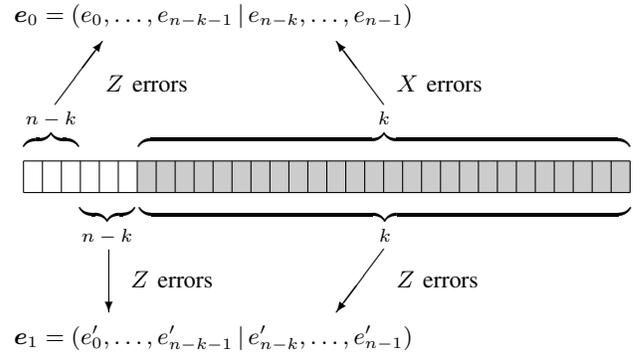
\begin{figure}
\centering
\begin{picture}(10,0.7)
\put(0,0){{\small $\boldsymbol{e}_0 = (e_0,\dots, e_{n-k-1}\, \vert\, e_{n-k},\dots, e_{n-1})$}}
\end{picture}

\begin{picture}(9.7,1.5)
\put(0.5,0.2){\vector(3,4){0.75}}
\put(1.3,0.4){{\small $Z$ errors}}

\put(5.7,0.2){\vector(-3,4){0.75}}
\put(5.9,0.4){{\small $X$ errors}}

\put(0,-0.8){$\overbrace{\phantom{lllll0}}^\text{$n-k$}\phantom{\underbrace{\phantom{lllll}}_\text{$n-k$}}
\overbrace{\phantom{lllllllllllllllllllllllllllllllllllllllllll000000llllll}}^\text{$k$}$}

\color[rgb]{0.8,0.8,0.8}
\put(1.8,-1.2){\rule{2.52mm}{4.2mm}}
\put(2.1,-1.2){\rule{2.52mm}{4.2mm}}
\put(2.4,-1.2){\rule{2.52mm}{4.2mm}}
\put(2.7,-1.2){\rule{2.52mm}{4.2mm}}
\put(3,-1.2){\rule{2.52mm}{4.2mm}}
\put(3.3,-1.2){\rule{2.52mm}{4.2mm}}
\put(3.6,-1.2){\rule{2.52mm}{4.2mm}}
\put(3.9,-1.2){\rule{2.52mm}{4.2mm}}
\put(4.2,-1.2){\rule{2.52mm}{4.2mm}}
\put(4.5,-1.2){\rule{2.52mm}{4.2mm}}
\put(4.8,-1.2){\rule{2.52mm}{4.2mm}}
\put(5.1,-1.2){\rule{2.52mm}{4.2mm}}
\put(5.4,-1.2){\rule{2.52mm}{4.2mm}}
\put(5.7,-1.2){\rule{2.52mm}{4.2mm}}
\put(6,-1.2){\rule{2.52mm}{4.2mm}}
\put(6.3,-1.2){\rule{2.52mm}{4.2mm}}
\put(6.6,-1.2){\rule{2.52mm}{4.2mm}}
\put(6.9,-1.2){\rule{2.52mm}{4.2mm}}
\put(7.2,-1.2){\rule{2.52mm}{4.2mm}}
\put(7.5,-1.2){\rule{2.52mm}{4.2mm}}
\put(7.8,-1.2){\rule{2.52mm}{4.2mm}}
\put(8.1,-1.2){\rule{2.52mm}{4.2mm}}
\put(8.4,-1.2){\rule{2.52mm}{4.2mm}}
\put(8.7,-1.2){\rule{2.52mm}{4.2mm}}
\put(9,-1.2){\rule{2.52mm}{4.2mm}}
\put(9.3,-1.2){\rule{2.52mm}{4.2mm}}
\color{black}

\put(0,-1.2){\line(1,0){9.6}}
\put(0,-0.7){\line(1,0){9.6}}

\put(0,-1.2){\line(0,1){0.5}}
\put(0.3,-1.2){\line(0,1){0.5}}
\put(0.6,-1.2){\line(0,1){0.5}}
\put(0.9,-1.2){\line(0,1){0.5}}
\put(1.2,-1.2){\line(0,1){0.5}}
\put(1.5,-1.2){\line(0,1){0.5}}
\put(1.8,-1.2){\line(0,1){0.5}}
\put(2.1,-1.2){\line(0,1){0.5}}
\put(2.4,-1.2){\line(0,1){0.5}}
\put(2.7,-1.2){\line(0,1){0.5}}
\put(3,-1.2){\line(0,1){0.5}}
\put(3.3,-1.2){\line(0,1){0.5}}
\put(3.6,-1.2){\line(0,1){0.5}}
\put(3.9,-1.2){\line(0,1){0.5}}
\put(4.2,-1.2){\line(0,1){0.5}}
\put(4.5,-1.2){\line(0,1){0.5}}
\put(4.8,-1.2){\line(0,1){0.5}}
\put(5.1,-1.2){\line(0,1){0.5}}
\put(5.4,-1.2){\line(0,1){0.5}}
\put(5.7,-1.2){\line(0,1){0.5}}
\put(6,-1.2){\line(0,1){0.5}}
\put(6.3,-1.2){\line(0,1){0.5}}
\put(6.6,-1.2){\line(0,1){0.5}}
\put(6.9,-1.2){\line(0,1){0.5}}
\put(7.2,-1.2){\line(0,1){0.5}}
\put(7.5,-1.2){\line(0,1){0.5}}
\put(7.8,-1.2){\line(0,1){0.5}}
\put(8.1,-1.2){\line(0,1){0.5}}
\put(8.4,-1.2){\line(0,1){0.5}}
\put(8.7,-1.2){\line(0,1){0.5}}
\put(9,-1.2){\line(0,1){0.5}}
\put(9.3,-1.2){\line(0,1){0.5}}
\put(9.6,-1.2){\line(0,1){0.5}}

\put(0,-1.4){$\phantom{\overbrace{\phantom{lll00}}_\text{$n-k$}}\underbrace{\phantom{lllll0}}_\text{$n-k$}
\underbrace{\phantom{lllllllllllllllllllllllllllllllllllllllllll000000llllll}}_\text{$k$}$}

\put(5.7,-2.1){\vector(-3,-4){0.75}}
\put(5.9,-2.7){{\small $Z$ errors}}
\put(1.35,-2.1){\vector(0,-1){1}}
\put(1.7,-2.7){{\small $Z$ errors}}
\end{picture}

\begin{picture}(10,3.6)
\put(0,0){{\small $\boldsymbol{e}_1 = (e'_0,\dots, e'_{n-k-1}\, \vert\, e'_{n-k},\dots, e'_{n-1})$}}
\end{picture}
\caption{Correspondence of quantum errors to error vectors. $\vert$
The white boxes represent the $2(n-k)$ less noisy qubits that may experience only phase errors.
The gray boxes are the $k$ noisy qubits that may suffer from bit and/or phase errors.
The first $n-k$ bits of $\boldsymbol{e}_0$ and the first $n-k$ bits of $\boldsymbol{e}_1$ correspond to
whether phase errors occurred on the $2(n-k)$ less noisy qubits.
The remaining $k$ bits of $\boldsymbol{e}_0$ indicate whether bit errors occurred on the $k$ noisy qubits
while the remaining $k$ bits of $\boldsymbol{e}_1$ correspond to phase errors on these noisy qubits.}\label{figcorrespondence}
\end{figure}
The point of Theorem \ref{lessnoisylemma} is that because $H$ is a parity-check matrix of a linear code of minimum distance $d$,
we can correctly infer $\boldsymbol{e}_0$ and $\boldsymbol{e}_1$ from
the syndromes $\boldsymbol{s}_0$ and $\boldsymbol{s}_1$,
which are $H\boldsymbol{e}_0^T$ and $H\boldsymbol{e}_1^T$ respectively,
if the weights of $\boldsymbol{e}_0$ and $\boldsymbol{e}_1$ are both less than or equal to $\lfloor\frac{d-1}{2}\rfloor$.
This implies that, by the definition of $\boldsymbol{e}_0$ and $\boldsymbol{e}_1$, the positions of all bit errors and phase errors can be identified.
Thus, the errors can be corrected if the number of physical qubits that suffer bit errors, phase errors or both is at most $\lfloor\frac{d-1}{2}\rfloor$.

While it is straightforward to derive Theorem \ref{lessnoisylemma} from the results already presented in \cite{Fujiwara:2013c},
for completeness, we give a formal proof in the remainder of this subsection.

For a unitary operator $U$ and a $v$-dimensional vector $\boldsymbol{a} = (a_0, \dots, a_{v-1}) \in \mathbb{F}_2^v$,
define $U^{\boldsymbol{a}}$ as the $v$-fold tensor product $O_0 \otimes \dots\otimes O_{v-1}$, where
$O_i = U$ if $a_i = 1$ and $O_i$ is the identity operator otherwise.

Take a linear $[n,k,d]$ code with a parity-check matrix $H$ in standard form
\[H = \left[\begin{array}{cc}I & A\end{array}\right]\]
for some $(n-k) \times k$ matrix $A$ over $\mathbb{F}_2$, where $I$ is the $(n-k) \times (n-k)$ identity matrix.
The $Z$-\textit{information check matrix} $H_Z$ and $X$-\textit{information check matrix} $H_X$ of $H$ are the $2(n-k) \times k$ matrices
\[H_Z = \left[\begin{array}{c}
A\\
0
\end{array}\right]\]
and
\[H_X = \left[\begin{array}{c}
0\\
A
\end{array}\right]\]
respectively.
Simply put, $H_Z$ and $H_X$ are matrices composed of the $(n-k) \times k$ all-zero matrix
and the columns of the parity-check matrix $H$ that correspond to the information bits.

Let $\ket{0}^{\otimes 2(n-k)}_X$ be $2(n-k)$ qubits in the joint $+1$ eigenstate of $X^{\otimes 2(n-k)}$.
Without loss of generality, we assume that $\ket{0}_X = \frac{\ket{0}+\ket{1}}{\sqrt{2}}$ and that $\ket{1}_X = \frac{\ket{0}-\ket{1}}{\sqrt{2}}$,
where $\ket{0}$ and $\ket{1}$ are the computational basis.

\begin{lemma}[\cite{Fujiwara:2013c}]\label{syndromelemma}
Assume that there exits a linear code of length $n$ and dimension $k$ with a parity-check matrix $H$ in standard form.
Define
\[Q = \sum_{\mu \in \mathbb{F}_2^{2(n-k)}}\ket{\mu}\bra{\mu}\otimes X^{\mu H_X}Z^{\mu H_Z}\]
and $Q^{\dag}$ as its complex conjugate, where $H_Z$ and $H_X$ are the $Z$- and $X$-information check matrices of $H$.
Take a pair $\boldsymbol{e}_X, \boldsymbol{e}_Z \in \mathbb{F}_2^{2n-k}$ of arbitrary $(2n-k)$-dimensional vectors.
Define ${\boldsymbol{e}_X}_l$ and ${\boldsymbol{e}_X}_r$ as
the first $2(n-k)$ and the remaining $k$ bits of ${\boldsymbol{e}_X}$ respectively so that $\boldsymbol{e}_X = ({\boldsymbol{e}_X}_l, {\boldsymbol{e}_X}_r)$.
Define similarly $\boldsymbol{e}_Z = ({{\boldsymbol{e}_Z}_l}_0, {{\boldsymbol{e}_Z}_l}_1, {\boldsymbol{e}_Z}_r)$,
where ${{\boldsymbol{e}_Z}_l}_0$, ${{\boldsymbol{e}_Z}_l}_1$, and ${\boldsymbol{e}_Z}_r$ are
the first $n-k$, the next $n-k$, and the last $k$ bits of ${\boldsymbol{e}_Z}$ respectively.
Let $\boldsymbol{e}_0 = ({{\boldsymbol{e}_Z}_l}_0, {\boldsymbol{e}_X}_r)$
and $\boldsymbol{e}_1 = ({{\boldsymbol{e}_Z}_l}_1, {\boldsymbol{e}_Z}_r)$.
Assume that ${\boldsymbol{e}_X}_l = 0$.
For arbitrary $k$ qubit state $\ket{\psi}$,
\begin{align}
Q^{\dag}&X^{\boldsymbol{e}_X}Z^{\boldsymbol{e}_Z}Q\ket{0}^{\otimes 2(n-k)}_X\ket{\psi}\nonumber\\
&= \ket{\left(H{\boldsymbol{e}_0}^T, H{\boldsymbol{e}_1}^T\right)}_X\label{encdeceq}
\otimes X^{{\boldsymbol{e}_X}_r}Z^{{\boldsymbol{e}_Z}_r}\ket{\psi}.
\end{align}
\end{lemma}

Theorem \ref{lessnoisylemma} immediately follows from the above lemma.

\begin{IEEEproof}[Proof of Theorem \ref{lessnoisylemma}]
Regard the arbitrary $k$ qubit state $\ket{\psi}$, unitary operator $Q$, and complex conjugate $Q^{\dag}$ in Lemma \ref{syndromelemma}
as the original $k$-qubit information which is to be encoded, an encoding operator, and a decoding operator respectively.
Assume that the supports $\operatorname{supp}(\boldsymbol{e}_X), \operatorname{supp}(\boldsymbol{e}_Z)$
of the pair $\boldsymbol{e}_X, \boldsymbol{e}_Z$ of arbitrary $(2n-k)$-dimensional vectors
represent the positions of $X$ errors and $Z$ errors introduced by a quantum channel respectively such that
an $X$ error occurred on the $i$th physical qubit if and only if $i \in \operatorname{supp}(\boldsymbol{e}_X)$ and such that
a $Z$ error occurred on the $i$th physical qubit if and only if $i \in \operatorname{supp}(\boldsymbol{e}_Z)$.
With these assumptions, the other assumption that ${\boldsymbol{e}_X}_l = 0$ made in Lemma \ref{syndromelemma} corresponds to
the condition that a fixed $2(n-k)$ physical qubits are only subject to phase errors.
Measuring the $2(n-k)$ ancilla qubits on the right-hand side of Equation (\ref{encdeceq}) in the Hadamard rotated basis
gives a $2k$-dimensional vector $\boldsymbol{s} \in \mathbb{F}_2^{2k}$ of which
the first half is the $k$-dimensional vector $\boldsymbol{s}_0 = H{\boldsymbol{e}_0}^T$
and the second half of which is $\boldsymbol{s}_1 = H{\boldsymbol{e}_1}^T$.
The proof is complete.
\end{IEEEproof}

Clearly, if $\left\vert \text{supp}(\boldsymbol{e}_0)\right\vert, \left\vert \text{supp}(\boldsymbol{e}_1)\right\vert \leq  \left\lfloor\frac{d-1}{2}\right\rfloor$,
the retrieved classical information in the form of a pair of $k$-dimensional vectors $H{\boldsymbol{e}_0}^T, H{\boldsymbol{e}_1}^T$
uniquely identifies the locations of nonzero bits in $\boldsymbol{e}_0$ and $\boldsymbol{e}_1$ as in standard syndrome decoding for linear codes.
The assumption that the number of quantum errors is $\lfloor\frac{d-1}{2}\rfloor$ or less,
which means $\left\vert \text{supp}(\boldsymbol{e}_X) \cup \text{supp}(\boldsymbol{e}_Z) \right\vert \leq \left\lfloor\frac{d-1}{2}\right\rfloor$,
implies that both $\left\vert \text{supp}(\boldsymbol{e}_0)\right\vert$ and $\left\vert \text{supp}(\boldsymbol{e}_1)\right\vert$
are less than or equal to $\left\lfloor\frac{d-1}{2}\right\rfloor$.
Trivially, once $\boldsymbol{e}_0$ and $\boldsymbol{e}_1$ are correctly inferred,
the two vectors $\boldsymbol{e}_X$ and $\boldsymbol{e}_Z$ that specify the positions of bit errors and phase errors can be fully reconstructed.

\subsection{Entanglement Assistance}\label{entanglement}
Here we review necessary basic facts on entanglement-assisted quantum error-correcting codes.
We follow the method used in \cite{Fujiwara:2010e,Hsieh:2011} for constructing quantum LDPC codes
through the entanglement-assisted analogue of the Calderbank-Shor-Steane (CSS) construction \cite{Calderbank:1996,Steane:1996}.
Similar to the previous method that uses less noisy qubits,
this entanglement-assisted method allows for extracting the information about what type of quantum error occurred on which qubit
by simply treating a pair of binary vectors as error syndromes.

An $[[n,k;c]]$ \textit{entanglement-assisted quantum error-correcting code} is an error correction scheme
that encodes $k$ logical qubits into $n$ physical qubits with the help of $c$ ebits.
The $c$ ebits are sent through a noiseless channel.
When importing a linear code with a parity-check matrix $H$,
the required number $c$ of noiseless qubits is exactly the $2$-rank of the matrix $HH^T$ over $\mathbb{F}_2$ \cite{Wilde:2008}.
Because we only consider ranks over $\mathbb{F}_2$,
in what follows we simply write $\operatorname{rank}(A)$ to mean the $2$-rank of a given matrix $A$.

The following is a straightforward consequence of the CSS construction for entanglement-assisted quantum error-correcting codes.
\begin{theorem}[\cite{Hsieh:2011}]\label{eaprop}
Assume that $n$ physical qubits are sent through a noisy quantum channel.
Define $\boldsymbol{e}_X = (e_0,\dots,e_{n-1}) \in \mathbb{F}_2^n$ to be the $n$-dimensional vector representing the positions of bit errors
such that $e_i = 1$ if a bit error occurred on the $i$th qubit and $e_i = 0$ otherwise.
Define also  $\boldsymbol{e}_Z = (e'_0,\dots,e'_{n-1}) \in \mathbb{F}_2^n$ to be the $n$-dimensional vector representing the positions of phase errors
such that $e'_i = 1$ if a phase error occurred on the $i$th qubit and $e'_i = 0$ otherwise.
Let $H$ be a parity-check matrix of a linear $[n,k,d]$ code.
There exists an $[[n,2k-n+\operatorname{rank}(HH^T); \operatorname{rank}(HH^T)]]$ entanglement-assisted quantum error-correcting code
that allows for retrieving classical information about quantum errors
in the form of a pair $\boldsymbol{s}_X, \boldsymbol{s}_Z \in \mathbb{F}_2^{n-k}$ of $(n-k)$-dimensional vectors
such that $\boldsymbol{s}_X = H\boldsymbol{e}_X^T$ and $\boldsymbol{s}_Z = H\boldsymbol{e}_Z^T$.
\end{theorem}

Because $H$ is a parity-check matrix of a linear $[n,k,d]$ code,
it is straightforward to see that if the number of bit errors and that of phase errors are both less than or equal to $\lfloor\frac{d-1}{2}\rfloor$,
the qubits on which bit errors occurred and the ones on which phase errors occurred can be identified.
Note that unlike in Theorem \ref{lessnoisylemma},
we do not require $H$ to be in standard form or of full rank in Theorem \ref{eaprop}.
Instead, typical and realistic assumptions require that the $2$-rank $\operatorname{rank}(HH^T)$ be very small
because it is the number of ebits we need to engineer extremely accurately and protect perfectly.
The most extreme case is when $\operatorname{rank}(HH^T) = 0$, where Theorem \ref{eaprop} reduces to the CSS construction in its original form.
Entanglement assistance takes place when $\operatorname{rank}(HH^T) \geq 1$.

\section{Assisted Quantum LDPC Codes}\label{LDPC}
In this section we study the desirable structures of parity-check matrices
for use in high-rate quantum error correction assisted by less noisy qubits or error-free ebits.
We make the conservative assumption that the receiver has no knowledge of possible correlations between bit errors and phase errors
so that the decoder approximates the quantum channel by two binary symmetric channels,
one of which introduces the operator $X$ independently on each physical qubit with probability $p_x$
and the other of which make the operator $Z$ act independently on each physical qubit with probability $p_z$.
Thus, in the case of codes assisted by less noisy auxiliary qubits,
the receiver employs two separate decoders for a linear code to infer $\boldsymbol{e}_0$ and $\boldsymbol{e}_1$ in Theorem \ref{lessnoisylemma}
under the condition that
the parity-check matrix $H$ and two binary vectors $\boldsymbol{s}_0 = H\boldsymbol{e}_0^T$, $\boldsymbol{s}_1 = H\boldsymbol{e}_1^T$ are given.
For entanglement-assisted quantum error-correcting codes,
two separated decoders are used to infer $\boldsymbol{e}_X$ and $\boldsymbol{e}_Z$ in Theorem \ref{eaprop}
from two binary vectors $\boldsymbol{s}_X = H\boldsymbol{e}_X^T, \boldsymbol{s}_Z = H\boldsymbol{e}_Z^T$ and the parity-check matrix $H$.
In both cases, the receiver employs the sum-product algorithm for inference \cite{MacKay:2003}.

It is notable that if the receiver has some knowledge of correlations between bit errors and phase errors,
this information can be incorporated into the decoding algorithm with an increase in decoding complexity
by carefully implementing a quantum analogue of belief propagation \cite{Poulin:2008,Leifer:2008}.
In fact, significant improvements in error correction performance have been reported in a very optimistic situation
where the receiver has perfect knowledge of a channel with a very strong correlation due to depolarizing noise \cite{MacKay:2004,Wang:2012,Maurice:2013}.
Compared to this ideal assumption, our setting assumes a smaller amount of exploitable information about the channel.
This allows us to give a conservative estimate on error correction performance as a likely lower bound for various situations and
avoid the risk of relying on unrealistically accurate knowledge of how quantum errors manifest on actual hardware.
For a discussion on how the decoder may be able to gain channel knowledge in practice for error correction purposes,
the interested reader is referred to \cite{Fujiwara:2014}.

We divide the remainder of this section into three subsections.
Section \ref{designs} provides the definitions of combinatorial designs we take advantage of for designing codes throughout this paper. 
In Section \ref{ldpcphasesec} we study parity-check matrices suitable
for use as quantum LDPC codes assisted by less noisy auxiliary qubits.
Desirable parity-check matrices for entanglement-assisted quantum LDPC codes are investigated in Section \ref{ldpcebitsec}.

\subsection{Combinatorial Designs}\label{designs}
Let $K$ be a subset of positive integers.
A \textit{pairwise balanced design} of \textit{order} $v$ and \textit{index} $1$ with \textit{block sizes} from $K$, denoted by PBD$(v,K,1)$,
is an ordered pair $(V, {\mathcal B})$, where $V$ is a nonempty finite set of $v$ elements,
called \textit{points}, and ${\mathcal B}$ is a set of subsets of $V$, called \textit{blocks}, that satisfies the following two conditions:
\begin{enumerate}
	\item[(i)] each unordered pair of distinct elements of $V$ appears in exactly one block of ${\mathcal B}$, 
        \item[(ii)] for every $B \in {\mathcal B}$ the cardinality $|B| \in K$.
\end{enumerate}
When $K$ is a singleton $\{\mu\}$, the PBD is a \textit{Steiner} $2$-\textit{design} of \textit{order} $v$
and \textit{block size} $\mu$, and is denoted by $S(2,\mu,v)$.
A simple counting argument shows that the number of blocks in an $S(2,\mu,v)$ is exactly $\frac{v(v-1)}{\mu(\mu-1)}$.
A PBD of order $v$ is \textit{trivial} if it has no blocks or consists of only one block of size $v$.
The trivial PBD with no blocks necessarily has only one point.

Define $\alpha(K) = \gcd\{\mu-1 \mid \mu \in K\}$ and $\beta(K) = \gcd\{\mu(\mu-1) \mid \mu \in K\}$.
Necessary conditions for the existence of a PBD$(v,K,1)$ are $v-1 \equiv 0 \pmod{\alpha(K)}$
and $v(v-1) \equiv 0 \pmod{\beta(K)}$ \cite{Beth:1999}. These conditions are known to be asymptotically sufficient.
\begin{theorem}[Wilson \cite{Wilson:1972d}]\label{th:existencePBD}
There exists a constant $v_K$ such that
for every $v > v_K$ satisfying $v-1 \equiv 0 \pmod{\alpha(K)}$ and $v(v-1) \equiv 0 \pmod{\beta(K)}$
there exists a \textup{PBD}$(v,K,1)$.
\end{theorem}

An \textit{incidence matrix} of a PBD $(V,{\mathcal B})$ with $\vert V \vert = v$ and $\vert \mathcal{B} \vert = b$
is a binary $v \times b$ matrix $H = (h_{i,j})$ with 
rows indexed by points, columns indexed by blocks, and $h_{i,j}=1$ if the $i$th point is contained in the $j$th block, and $h_{i,j}=0$ otherwise.

It is known that incidence matrices of PBDs of index $1$ are generally good candidates of parity-check matrices of LDPC codes
for high speed information transmission because of their good error tolerance at relatively short lengths
\cite{Vasic:2004,Ammar:2004,Johnson:2003,Kou:2001}.
Our goal in the following two subsections is to identify and give explicit constructions for particularly promising classes of PBDs
whose incidence matrices may be used as parity-check matrices for assisted quantum error correction.

\subsection{Parity-Check Matrices for Phase Error Qubit Assistance}\label{ldpcphasesec}
The explicit restriction on the structure of a parity-check matrix $H$ of a linear $[n,k,d]$ code in Theorem \ref{lessnoisylemma} is
that it must be in standard form
\[H = \left[\begin{array}{cc}I & A\end{array}\right]\]
for some $(n-k) \times k$ matrix $A$ over $\mathbb{F}_2$, where $I$ is the $(n-k) \times (n-k)$ identity matrix.
As we will see later in this subsection, incidence matrices of PBDs of index $1$ may be seen as parity-check matrices
that give the largest possible information rates among all possible $H$ avoiding certain undesirable structures for the standard sum-product algorithm.
Because our goal is to construct promising parity-check matrices of LDPC codes of extremely high rate,
here we would like $H$ as a whole to form an incidence matrix of a PBD of index $1$.
The following proposition allows us to only consider the part $A$ in this regard.

\begin{proposition}\label{propPBD}
Let $H = \left[\begin{array}{cc}I & A\end{array}\right]$ be a parity-check matrix of a linear code of length $n$, dimension $k$,
and minimum distance larger than $2$ in standard form.
$H$ is an incidence matrix of a PBD of index $1$ if and only if the $(n-k) \times k$ matrix $A$ is an incidence matrix of a PBD of index $1$
containing no block of size $1$.
\end{proposition}
\begin{IEEEproof}
Assume that $H$ is a parity-check matrix in standard form that forms an incidence matrix of a PBD of index $1$.
Because the condition on the minimum distance dictates that no pair of columns be identical,
the PBD contains exactly $n-k$ blocks of size $1$, which correspond to the $n-k$ columns of weight $1$ in $H$.
Because these blocks do not contribute to the number of each pair of points appearing in blocks,
deleting the corresponding $(n-k)\times(n-k)$ identity matrix $I$ leaves $(n-k) \times k$ matrix $A$, where
every pair of rows have exactly one column in which both rows have $1$.
Indexing rows by points and columns by blocks, $A$ forms an incidence matrix of a PBD of index $1$.
Because every column of weight $1$ is deleted from $H$, this PBD does not have a singleton as a block.
Conversely, because a block of size $1$ does not have a pair of points,
combining the $(n-k)\times(n-k)$ identity matrix $I$ and an incidence matrix $A$ of a PBD$(n-k,K,1)$ with $1 \not\in K$
gives an incidence matrix $H = \left[\begin{array}{cc}I & A\end{array}\right]$ of a PBD of index $1$.
\end{IEEEproof}

In view of the above proposition, we would like to find incidence matrices $A$ of PBDs without blocks of size $1$ that do not contain or produce harmful structures
when combined with the identity matrices to obtain valid parity-check matrices $H = \left[\begin{array}{cc}I & A\end{array}\right]$ for Theorem \ref{lessnoisylemma}.
While it is generally a very difficult open problem to exactly determine relative harmfulness of each substructure of a parity-check matrix for
the sum-product algorithm over a binary symmetric channel,
there are known structures that have theoretically or empirically been shown to be undesirable (see \cite{Nguyen:2012} and references therein).
We first consider a few of the more harmful structures.

The \textit{Tanner graph} of an $m \times n$ parity-check matrix $H$
is the bipartite graph consisting of $n$ \textit{bit vertices} indexed by bits of the corresponding code
and $m$ \textit{parity-check vertices} indexed by parity-check equations defined by $H$,
where an edge joins a bit vertex to a parity-check vertex if the bit is involved in the corresponding parity-check equation.
An $l$-\textit{cycle} in a graph is a sequence of $l+1$ connected vertices
which starts and ends at the same vertex in the graph and contains no other vertices more than once.
Clearly, a $4$-cycle in a Tanner graph is equivalent to a $2 \times 2$ all-one submatrix in a parity-check matrix.
A $6$-cycle is a $3 \times 3$ submatrix in which each row and column has exactly two ones.
The \textit{girth} of a parity-check matrix is the length of a shortest cycle in the corresponding Tanner graph.
Since a Tanner graph is bipartite, its girth is always even.
When it is clear from context which parity-check matrix is considered,
we may speak of the ``girth of an LDPC code.''
It is known that very short cycles tend to be harmful when the sum-product algorithm is employed.
In particular, $4$-cycles have a very noticeable negative effect on the error correction performance of the sum-product algorithm \cite{Johnson:2010}.
For this reason, we would like the girth of a parity-check matrix to be strictly larger than $4$.

Because an LDPC code is a linear code equipped with a particular decoding algorithm,
the minimum distance also plays a role.
While the sum-product algorithm is generally less sensitive to the minimum distance than other simple decoding methods,
this fundamental parameter is especially important to a code of very high rate because its very large dimension dictates that
the minimum distance be small compared to the length.
The following proposition concerns with the number of short cycles and minimum distances of parity-check matrices
based on incidence matrices of PBDs together with columns of weight $1$.

\begin{proposition}\label{cycledistancePBD}
Let $A$ be an $(n-k)\times k$ incidence matrix of a nontrivial PBD$(n-k,K,1)$ with $1 \not\in K$.
Then the binary matrix $H = \left[\begin{array}{cc}I & A\end{array}\right]$
is a parity-check matrix of a linear $[n,k,d]$ code in standard form whose girth is $6$ and minimum distance $d=1+\min\{\mu \mid \mu \in K\}$.
\end{proposition}
\begin{IEEEproof}
We first prove that the parity-check matrix $H$ is of girth $6$.
Because no pair of points appear twice in a PBD of index $1$,
there exits no $2 \times 2$ all-one submatrix in $A$.
Hence, the girth of $A$ is at least $6$.
Take an arbitrary column $\boldsymbol{c_1}$ of $A$.
Write the block $B_1$ which corresponds to $\boldsymbol{c_1}$ as $\{v_1, \dots, v_{|B_1|}\}$.
Because the PBD is nontrivial, every row of $A$ has at least two ones.
Thus, there exists a column $\boldsymbol{c_2}$
which corresponds to another block $B_2 = \{v_1, v_{|B_1|+1},\dots, v_{|B_1|+|B_2|-1}\}$,
where $v_i \not= v_j$ for any $i$ and $j$, $i \not= j$.
Take the column $\boldsymbol{c_3}$
representing the block $B_3$ that contains the pair $\{v_2, v_{{|B_1|}+1}\}$.
The three columns $\boldsymbol{c_1}$, $\boldsymbol{c_2}$, and $\boldsymbol{c_3}$ induce a $6$-cycle,
which implies that the girth of $A$ is exactly $6$.
Since joining the identity matrix $I$ does not introduce $4$-cycles, the girth of $H$ is exactly $6$.

Next we show that the linear code is of minimum distance $1 + \min\{\mu \mid \mu \in K\}$.
It suffices to show that a smallest set of linearly dependent columns in $H$ is of cardinality $1+\min\{\mu \mid \mu \in K\}$.
Because no pair of points appear twice in a PBD of index $1$, any set of linearly dependent columns that contains at least one column of $A$
is of cardinality at least $1+\min\{\mu \mid \mu \in K\}$. All columns in the identity matrix are linearly independent.
Thus, we only need to show that there exits a set of exactly $1+\min\{\mu \mid \mu \in K\}$ linearly dependent columns of $H$.
Take an arbitrary column $\boldsymbol{c}$ of $A$ whose weight is the smallest.
The identity matrix $I$ contains the set $S$ of columns of cardinality $\min\{\mu \mid \mu \in K\}$ such that
$S \cup \{\boldsymbol{c}\}$ forms a set of linearly dependent columns.
Because the weight  of $\boldsymbol{c}$ is the smallest among all columns of $A$,
the cardinality $\vert S \cup \{\boldsymbol{c}\}\vert$ is $1+\min\{\mu \mid \mu \in K\}$.
\end{IEEEproof}

As we stated earlier in this subsection, incidence matrices of PBDs of index $1$ are extreme in terms of information rates.
The following proposition shows that if a column weight distribution admits a PBD of index $1$,
it is impossible to obtain a parity-check matrix of girth $6$ or higher without using an incidence matrix of some PBD of index $1$.
\begin{proposition}\label{automaticallyPBD}
Let $H$ be a parity-check matrix that forms an incidence matrix of a nontrivial PBD of order $n-k$ and index $1$.
Any parity-check matrix of the same size, same column weight distribution, and same or higher girth as $H$
is an incidence matrix of a PBD of order $n-k$ and index $1$.
\end{proposition}
\begin{IEEEproof}
Let $H$ be an $(n-k)\times n$ parity-check matrix that forms an incidence matrix of a nontrivial PBD of index $1$.
Define $\boldsymbol{w}_c = (w_0, w_1, \dots, w_{n-1})$ to be the $n$-dimensional vector over nonnegative integers $\mathbb{N}_0$
such that the column $\boldsymbol{c}_i = (c_0, c_1, \dots, c_{n-k})$ of $H = (\boldsymbol{c}_0, \dots, \boldsymbol{c}_{n-1})$ contains exactly $w_i$ $1$'s.
Each entry of the vector $\boldsymbol{w}_c$ represents the weight of each column of $H$.
Take a parity-check matrix $H'$ of the same size, same column weight distribution, and same or higher girth as $H$.
As in the case of $H$, define $\boldsymbol{w}'_c = (w'_0, w'_1, \dots, w'_{n-1})$ to be the $n$-dimensional vector
such that the column $\boldsymbol{c}'_i = (c'_0, c'_1, \dots, c'_{n-k})$ of $H' = (\boldsymbol{c}'_0, \dots, \boldsymbol{c}'_{n-1})$ contains exactly $w'_i$ $1$'s.
Assume to the contrary that $H'$ is not an incidence matrix of a PBD of index $1$.
We prove that this leads to a contradiction.

As usual, we define $\binom{a}{b}$ to be $0$ when $0 < a < b$,
so that the binomial coefficient counts the number of ways to choose $b$ elements from a finite set of positive cardinality $a$.
Recall that $H$ is an incidence matrix of a PBD of index $1$ with $n-k$ points, which means that
each pair of points appears exactly once in blocks.
Hence, adding up the number of pairs in each block gives
\begin{align}\label{counting}
\sum_{i=0}^{n-1}\binom{w_i}{2} = \binom{n-k}{2}.
\end{align}
Note that Equation (\ref{counting}) only depends on $n$, $k$, and each value of $w_i$.
Because $H'$ has the same column weight distribution as that of $H$,
the vector $\boldsymbol{w}'_c$ is obtained by permuting the coordinates of $\boldsymbol{w}_c$.
Hence, we also have
\begin{align}\label{counting2}
\sum_{i=0}^{n-1}\binom{w'_i}{2} = \binom{n-k}{2}.
\end{align}

Now, the left-hand side of Equation (\ref{counting2}) can be interpreted as counting the number of $2 \times 1$ all-one submatrix in $H'$,
which implies that $H'$ has exactly $\binom{n-k}{2}$ $2 \times 1$ all-one submatrices.
If no pair of $2 \times 1$ all-one submatrices arises in the same pair of rows,
by indexing rows and columns by points and blocks respectively,
$H'$ forms an incidence matrix of a PBD of index $1$, a contradiction.
Thus, there is a $2\times2$ all-one submatrix in $H'$.
However, a $2\times2$ all-one submatrix gives rise to a $4$-cycle, contradicting the assumption that $H'$ is of girth $6$ or higher.
The proof is complete.
\end{IEEEproof}

One might hope for a higher rate without decreasing the girth or dimension by changing the number of parity-check equations.
Since increasing the number of rows decreases the dimension, we need to use fewer rows.
However, if we use a parity-check matrix with a smaller number $n-k-x$ of rows for some positive $x$,
because $\binom{n-k-x}{2} < \binom{n-k}{2}$, the resulting LDPC code necessarily contains a $4$-cycle.
Note that given the number of rows, the dimension of a parity-check matrix in standard form is determined by the number of information bits.
In other words, the longer the linear code is, the higher the information rate will be.
Therefore, if we impose some restriction on the column weight distribution such as that every column is of the same weight
or that the maximum column weight is $c$ for some positive constant $c$,
in order for a parity-check matrix to achieve the highest dimension among those that satisfy the given condition,
the sum of the number of $2 \times 1$ submatrices in each column as in the left-hand side of Equation (\ref{counting2}) should be as large as possible.
An incidence matrix of a PBD of index $1$ is an extremal example in that it achieves the upper bound $\binom{n-k}{2}$
for parity-check matrices that do not contain $4$-cycles.

An incidence matrix $A$ of a PBD$(n-k,K,1)$ gives an LDPC code of minimum distance $1+\min\{\mu \mid \mu \in K\}$ when combined with the identity matrix.
Hence, increasing the smallest block size improves the minimum distance.
However, because a block of size $x$ contains ${{x}\choose{2}}$ pairs, a block of larger size contains more pairs of points.
Since avoiding $4$-cycles while achieving the highest possible rate is equivalent to
packing as many different pairs of points as possible in a set of blocks while including no pair of points more than once,
increasing block sizes lowers the achievable information rate in general.
Hence, we consider the case when the column weights of the matrix $A$ are uniform.
This means that $K$ is a singleton $\{\mu\}$, that is, the corresponding PBD$(n-k,K,1)$ is a Steiner $2$-design $S(2,\mu,n-k)$.
As stated earlier in Section \ref{designs}, an $S(2,\mu,v)$ contains exactly $\frac{v(v-1)}{\mu(\mu-1)}$ blocks.
Thus, the corresponding code dimension can be quite large at a moderate length.
\begin{proposition}\label{linearS}
Let $A$ be an incidence matrix of an $S(2,\mu,v)$ and $I$ a $v \times v$ identity matrix.
A parity-check matrix $H = \left[\begin{array}{cc}I & A\end{array}\right]$ defines an LDPC code of
length $\frac{v(v-1)}{\mu(\mu-1)}+v$, dimension $\frac{v(v-1)}{\mu(\mu-1)}$, girth $6$, and minimum distance $\mu+1$.
\end{proposition}

As can be seen from the above proposition,
the rates of LDPC codes defined by incidence matrices of $S(2,\mu,v)$s become close to $1$ very quickly as $v$ tends to infinity.
Theorems \ref{binaryonlyZ} and \ref{lessnoisylemma} assure that the corresponding quantum error-correcting codes assisted by less noisy qubits
inherit this characteristic.
\begin{theorem}\label{quantumS}
Let $A$ be an incidence matrix of an $S(2,\mu,v)$ and $I$ a $v \times v$ identity matrix.
There exits a $[[\frac{v(v-1)}{\mu(\mu-1)}+2v,\frac{v(v-1)}{\mu(\mu-1)}]]$ quantum error-correcting code
that identifies the types and locations of quantum errors
through the LDPC code defined by the parity-check matrix $H = \left[\begin{array}{cc}I & A\end{array}\right]$
under the assumption that a fixed set of $2(n-k)$ physical qubits may experience phase errors but no bit errors.
\end{theorem}

One strategy to improve the error correction performance under the sum-product algorithm
is to decrease the number of structures that are responsible for dominating errors.
While joining the identity matrix and the incidence matrix $A$ of a Steiner $2$-design of block size $\mu$
always results in an LDPC code of minimum distance $\mu+1$,
it is desirable for the linear code defined by $A$ alone to have a larger minimum distance
because it eliminates dominating sources of errors to an extent.

It is trivial that the minimum distance of a linear code whose parity-check matrix forms an incidence matrix of an $S(2,\mu,v)$ is at least $\mu + 1$.
To investigate the minimum distances of linear codes based on Steiner $2$-designs further,
we define some combinatorial design theoretic notions.
A \textit{configuration} ${\mathcal C}$ in an $S(2,\mu,k)$, $(V,{\mathcal B})$,
is a subset ${\mathcal C} \subseteq {\mathcal B}$ of the block set.
The set of points appearing in at least one block of a configuration ${\mathcal C}$
is denoted by $V({\mathcal C})$.
Two configurations ${\mathcal C}$ and ${\mathcal C}'$ are \textit{isomorphic}
if there exists a bijection $\phi : V({\mathcal C}) \rightarrow V({\mathcal C}')$
such that for each block $B \in {\mathcal C}$,
the image $\phi(B)$ is a block in ${\mathcal C}'$.
When $|{\mathcal C}|=i$,
a configuration ${\mathcal C}$ is  called an \textit{$i$-configuration}.
A configuration ${\mathcal C}$ is  \textit{even}
if for every point $a$ appearing in ${\mathcal C}$
the number $|\{B \mid a \in B \in {\mathcal C}\}|$
of blocks containing $a$ is even.

The notion of minimum distance can be described in the language of combinatorial designs.
An $S(2,\mu,v)$ is  {\it $r$-even-free} if for every integer $i$ satisfying $1\leq i \leq r$ it contains no even $i$-configurations.
Because the minimum distance of a linear code is the size of a smallest linearly dependent set of columns in its parity-check matrix,
the minimum distance of a linear code based on an incidence matrix $A$ of a Steiner $2$-design is determined by its even-freeness.
\begin{proposition}\label{evenfree=min}
The minimum distance of a linear code whose parity-check matrix forms an incidence matrix of a Steiner $2$-design is $d$
if and only if the corresponding Steiner $2$-design is $(d-1)$-even-free but not $d$-even-free.
\end{proposition}

By definition every $r$-even-free $S(2,\mu,v)$, $r\geq 2$, is also $(r-1)$-even-free.
If $\mu$ is odd, a simple double counting argument shows that an $r$-even-free $S(2,\mu,v)$ with $r$ even is also $(r+1)$-even-free.
Because a Steiner $2$-design is a linear space in the sense of incidence geometry, every $S(2,\mu,v)$ is trivially $\mu$-even-free.

A nontrivial $S(2,\mu,v)$ may or may not be $(\mu+1)$-even-free.
For each $\mu\geq2$, an even $(\mu+1)$-configuration that may arise in $S(2,\mu,v)$s is unique up to isomorphism;
they are the dual of the complete graph on $\mu+1$ vertices.
For instance, for the case when $\mu = 3$, up to isomorphism, there exists only one possible even $4$-configuration, called the {\it Pasch} configuration.
It can be written by six points and four blocks:
\[\{\{a, b, c\}, \{a, d, e\}, \{f, b, d\}, \{f, c, e\}\}.\]
The unique possible even $(\mu+1)$-configurations for $\mu \geq 4$ are sometimes called the \textit{generalized Pasch} configurations
in the coding theory literature (see, for example, \cite{Vasic:2004,Gruner:2013}).
Since they are the smallest and unique, an $S(2,\mu,v)$ is $(\mu+1)$-even-free if and only if it contains no Pasch configurations for $\mu=3$
and no generalized Pasch configurations for $\mu \geq 4$.

A fairly tight bound on the maximum even-freeness of an $S(2,3,v)$ is available.
\begin{theorem}[\cite{Fujiwara:2010}]\label{no8evenfreeSTS}
There exists no nontrivial $8$-even-free $S(2,3,v)$.
\end{theorem}
Hence, by Proposition \ref{evenfree=min}, Theorem \ref{no8evenfreeSTS}, and the fact that every $S(2,3,v)$ is $3$-even-free,
we obtain bounds on the minimum distance.
\begin{theorem}\label{th:paraSTSEAQECC}
The minimum distance $d$ of a linear code whose parity-check matrix forms an incidence matrix of a nontrivial $S(2,3,v)$ satisfies
the inequalities $4 \leq d \leq 8$.
\end{theorem}

The problem of avoiding Pasch configurations has long been investigated in various contexts in discrete mathematics.
The fundamental question that asks which order $v$ admits an $S(2,3,v)$ avoiding Pasch configurations was settled in 2000 \cite{Grannell:2000}.
Note that such $S(2,3,v)$s are $4$-even-free and hence are automatically $5$-even-free due to their block size being odd number $3$.
\begin{theorem}[\cite{Grannell:2000}]\label{PaschSTS}
There exists a $5$-even-free $S(2,3,v)$ if and only if $v \equiv 1, 3 \pmod{6}$ except $v = 7, 13$.
\end{theorem}

While attaining $(\mu+1)$-even-freeness in the right portion $A$ of our parity-check matrix $H = \left[\begin{array}{cc}I & A\end{array}\right]$
is good enough to achieve the goal of reducing the number of codewords of the smallest weight,
if one wishes even higher even-freeness, it is required to construct an $S(2,3,v)$ that simultaneously avoids Pasch and two more even configurations,
namely the \textit{grids}
\[\{\{a,b,c\}, \{d,e,f\}, \{g,h,i\}, \{a,d,g\}, \{b,e,h\}, \{c,f,i\}\}\]
and \textit{double triangles}
\[\{\{a,b,c\},\{a,d,e\}, \{b,f,g\}, \{c,h,e\}, \{d,g,i\}, \{f,h,i\}\}.\]
Unfortunately, while there exist infinitely many $S(2,3,v)$s avoiding both Pasch and double triangle configurations \cite{Colbourn:2009},
no nontrivial examples avoiding grids, let alone $6$-even-free $S(2,3,v)$s, are known at the time of writing \cite{Furedi:2013}.
If a nontrivial $S(2,3,v)$ that simultaneously avoids the three even configurations exists,
it is automatically $7$-even-free and hence attains the upper bound given in Theorem \ref{th:paraSTSEAQECC}
on the minimum distance of the corresponding linear code.

It is tempting to prove similar theorems on the even-freeness of $S(2,\mu,v)$s for all $\mu \geq 4$.
Unfortunately, while it appears that in principle some of the analogous mathematical arguments likely work \cite{Beezer:2001},
it seems very difficult to obtain equally tight bounds and/or complete existence results for relatively high even-freeness for general block size $\mu$.
In fact, no nontrivial upper bounds seem to be known on the even-freeness of $S(2,\mu,v)$s with large $\mu$
or, equivalently, on the minimum distances of the corresponding LDPC codes in general.
To the best of the authors' knowledge,
the only useful and fairly general bound is the one for $S(2,\mu,v)$s with special automorphisms.
\begin{theorem}[\cite{Fujiwara:2012}]\label{boundabelian}
If an abelian group acts transitively on the points of a nontrivial $r$-even-free $S(2,\mu,v)$ with $v>\mu(\mu-1)+1$, then $r \leq 2\mu-1$.
\end{theorem}

The usefulness of the above bound lies in the fact that the kind of $S(2,\mu,v)$ that is easy to analyze
and likely has higher even-freeness than the trivial lower bound suggests
tends to possess the algebraic property considered in Theorem \ref{boundabelian}.
In fact, all known nontrivial $S(2,\mu,v)$s with the highest even-freeness for $\mu\geq4$ admit such abelian group actions
and achieve the upper bound given in Theorem \ref{boundabelian}.

The \textit{affine geometry} $\textup{AG}(m,q)$ of \textit{dimension} $m$ over $\mathbb{F}_q$ is defined as a finite geometry in which
the \textit{points} are the vectors in $\mathbb{F}_q^m$
and the $i$-\textit{dimensional affine subspaces} are the $i$-dimensional vector subspaces of $\mathbb{F}_q^m$ and their cosets.
The points and $1$-dimensional affine subspaces of $\textup{AG}(m,q)$ form the points and blocks of an $S(2,q,q^m)$ \cite{Beth:1999}.
Affine geometries provide an explicit construction for nontrivial $S(2,\mu,v)$s with the highest known even-freeness.
\begin{theorem}[\cite{Muller:2004b}]\label{affineerasure}
For any odd prime power $q$ and positive integer $m \geq 2$
the points and $1$-dimensional affine subspaces of $\textup{AG}(m,q)$ form a $(2q-1)$-even-free $S(2,q,q^m)$ which is not $2q$-even-free.
\end{theorem}

Affine geometries are not the only known nontrivial $S(2,\mu,v)$s that attain the upper bound given in Theorem \ref{boundabelian}.
The \textit{projective geometry} $\textup{PG}(m,q)$ of \textit{dimension} $m$ over $\mathbb{F}_q$
is a finite geometry whose \textit{points} and $i$-\textit{dimensional subspaces} are the $1$-dimensional vector subspaces
and the $(i+1)$-dimensional vector subspaces of $\mathbb{F}_q^{m+1}$ respectively.
The points and $1$-dimensional subspaces of $\textup{PG}(m,q)$ form the points and blocks of an $S(2,q+1,\frac{q^{m+1}-1}{q-1})$.
\begin{theorem}[\cite{Fujiwara:2010e}]\label{projectiveerasure}
For any odd prime power $q$ and positive integer $m \geq 3$
the points and $1$-dimensional subspaces of $\textup{PG}(m,q)$ form a $(2q+1)$-even-free $S(2,q+1,\frac{q^{m+1}-1}{q-1})$ which is not $(2q+2)$-even-free.
\end{theorem}

Note that because the rank of an incidence matrix of the $S(2,\mu,v)$ from $\textup{PG}(m,q)$ with $q$ odd is $v-1$ \cite{Frumkin:1990a},
the $S(2,q+1,q^2+q+1)$ forming $\textup{PG}(2,q)$ with $q$ odd vacuously achieves the highest possible even-freeness $q^2+q$.

Recently, the first author gave a combinatorial construction for $(\mu+1)$-even-free $S(2,\mu,v)$s \cite{Fujiwara:2012}.
The construction technique recursively combines a $(\mu+1)$-even-free $S(2,\mu,v)$ and another $(\mu+1)$-even-free $S(2,\mu,w)$
with a particular algebraic property by using a specially designed combinatorial matrix in order to generate a larger $(\mu+1)$-even-free $S(2,\mu,vw)$.
For the details of the construction, we refer the reader to the original article \cite{Fujiwara:2012}.
As far as the authors are aware, no constructions for $S(2,\mu,v)$s with even-freeness higher than or equal to $\mu+1$ are known except
the finite geometric and recursive ones.

From the viewpoint of quantum error correction assisted by less noisy qubits,
the highly even-free $S(2,\mu,v)$s based on projective and affine geometries have an additional appealing property.
As described in Section \ref{lessnoisy}, our auxiliary qubits are assumed to be engineered more reliably than the rest
so that only phase errors may occur.
Hence, it would not be too unnatural to assume that those phase errors that may still occur on these qubits
manifest less frequently than bit errors and phase errors on the other qubits.
By Equation (\ref{encdeceq}) of Lemma \ref{syndromelemma}, in the language of linear codes,
this slightly more optimistic assumption translates into the premise that
the probability that an error occurs on a fixed check bit, which corresponds to a column of the $(n-k)\times(n-k)$ identity matrix $I$
in $H = \left[\begin{array}{cc}I & A\end{array}\right]$, is smaller than that on a fixed information bit corresponding to a column of $A$.
The following theorem shows that highly even-free $S(2,\mu,v)$s from finite geometries can take advantage of this nonuniformity.
\begin{theorem}[\cite{Muller:2004b,Vandendriessche:2013}]\label{AGoddpointtheorem}
Let $q$ be an odd prime power and $m \geq 2$ an integer greater than or equal to $2$.
Define $(V,\mathcal{B})$ to be the $(2q-1)$-even-free $S(2,q,q^m)$ formed by the points and $1$-dimensional affine subspaces of $\textup{AG}(m,q)$.
For any nonempty configuration $\mathcal{C} \subset \mathcal{B}$ whose size is in the range $1 < \vert\mathcal{C}\vert \leq 2q-1$,
it holds that
\[\vert\mathcal{C}\vert+\operatorname{odd}(\mathcal{C}) \geq 2q,\]
where $\operatorname{odd}(\mathcal{C})$ is the number of points $v \in V$ contained in an exactly odd number of blocks in $\mathcal{C}$.
\end{theorem}
\begin{theorem}[\cite{Vandendriessche:2013}]\label{PGoddpointtheorem}
Let $q$ be an odd prime power and $m \geq 2$ an integer greater than or equal to $3$.
Define $(V,\mathcal{B})$ to be the $(2q+1)$-even-free $S(2,q+1,\frac{q^{m+1}-1}{q-1})$
formed by the points and $1$-dimensional subspaces of $\textup{PG}(m,q)$.
For any nonempty configuration $\mathcal{C} \subset \mathcal{B}$ whose size satisfies $1 < \vert\mathcal{C}\vert \leq 2q+1$,
it holds that
\[\vert\mathcal{C}\vert+\operatorname{odd}(\mathcal{C}) \geq 2q+2.\]
\end{theorem}

The same inequality $\vert\mathcal{C}\vert+\operatorname{odd}(\mathcal{C}) \geq 2q+2$ as in Theorem \ref{PGoddpointtheorem}
holds also for the $S(2,q+1,q^2+q+1)$ from $\textup{PG}(2,q)$ with $q$ odd \cite{Vandendriessche:2013}.
The point of the above theorems is that a linear code of minimum distance $\mu+1$
defined by a parity-check matrix $H = \left[\begin{array}{cc}I & A\end{array}\right]$
with $A$ being an incidence matrix of a finite geometric $S(2,\mu,v)$ would perform better if check bits suffer from errors much less likely than information bits.
This is because, in a sense, it is effectively of minimum distance $2\mu$ except that
there is only one type of small weight codeword, which is the one that consists of one information bit and the corresponding $\mu$ check bits.
An error of this kind involving check bits would be unlikely to occur
if the additional assumption that the more reliable qubits have a sufficiently smaller error probability is valid.

It is notable that this effect of almost doubled minimum distances would be favorable across many different decoding methods and algorithms.
In the case of iterative decoding, a nonempty set of bit vertices in a Tanner graph
that are not correct after $l$ iterations for all $l \geq l_c$ for some absolute constant $l_c$ is called a \textit{trapping set} \cite{Nguyen:2012}.
To improve the error floor and slope of the block error rate curve,
it is desirable for a parity-check matrix to avoid small trapping sets \cite{Ivkovic:2008}.
While the set of small trapping sets generally varies from algorithm to algorithm and notoriously difficult to identify,
codewords of very small weight are surely among them.
We will demonstrate the performance of $S(2,\mu,v)$s based on finite geometries in the context of quantum error correction
assisted by more reliable auxiliary qubits in Section \ref{sim} through simulations.

\subsection{Parity-Check Matrices for Noiseless Qubit Assistance}\label{ldpcebitsec}
We now turn our attention to parity-check matrices suitable to entanglement-assisted quantum LDPC codes.
In order to put our results in context and show how the two types of assisted quantum error correction are related,
we quote the most relevant results that can be found in \cite{Hsieh:2011,Fujiwara:2010e,Fujiwara:2013b}
and also present some useful results that are known in combinatorics but are apparently not found in the quantum coding theory literature.
We then give a new method for finding promising high-rate entanglement-assisted quantum LDPC codes at the end of this section.

For entanglement assistance, our parity-check matrices do not need to be in standard form,
which was mandatory in the case of quantum error correction assisted by qubits with possible phase errors but no bit errors.
The unique requirement in the case of entanglement assistance is that, as mentioned earlier in Section \ref{entanglement},
the $2$-rank $\operatorname{rank}(HH^T)$ of the product of our parity-check matrix $H$ and its transpose must be kept small.
Thus, in view of Theorem \ref{eaprop} and the discussions given in the previous subsection on error correction performance of incidence matrices of PBDs
and their extremely high rates, our interest is in those PBDs that have high even-freeness and very small $2$-ranks $\operatorname{rank}(HH^T)$.

To keep our discussion succinct and directly take advantage of most of the material given in the previous subsections,
we focus mostly on a class of LDPC codes
in which the column weights and the row weights of parity-check matrices are both uniform, that is, \textit{regular} LDPC codes.
It is straightforward to see that an incidence matrix of an $S(2,\mu,v)$ is of constant column weight $\mu$ and constant row weight $\frac{v-1}{\mu-1}$,
providing a parity-check matrix of a regular LDPC code.
Since an $S(2,\mu,v)$ contains exactly $\frac{v(v-1)}{\mu(\mu-1)}$ blocks, by Theorem \ref{eaprop} and Proposition \ref{evenfree=min},
the parameters of the corresponding entanglement-assisted quantum LDPC code are as follows.
\begin{theorem}[\cite{Fujiwara:2010e}]\label{paraealdpc}
An incidence matrix $H$ of an $r$-even-free $S(2,\mu,v)$ that is not $(r+1)$-even-free gives an entanglement-assisted quantum LDPC code
of length $\frac{v(v-1)}{\mu(\mu-1)}$ and dimension $\frac{v(v-1)}{\mu(\mu-1)}-2\operatorname{rank}(H)+\operatorname{rank}(HH^T)$
that requires $\operatorname{rank}(HH^T)$ ebits
for quantum error correction through the LDPC code
of the same length, dimension $\frac{v(v-1)}{\mu(\mu-1)}-\operatorname{rank}(H)$, girth $6$, and minimum distance $r+1$
formed by $H$ as its parity-check matrix.
\end{theorem}

As shown in Proposition \ref{evenfree=min}, the minimum distance of the LDPC code from an incidence matrix of an $S(2,\mu,v)$ is dictated by its even-freeness.
Hence, the bounds and constructions for highly even-free Steiner $2$-designs given
in Theorems \ref{th:paraSTSEAQECC}, \ref{PaschSTS}, \ref{boundabelian}, \ref{affineerasure} and \ref{projectiveerasure} are fully and directly applicable here.

Unlike less noisy qubit assistance, however, the dimension of an entanglement-assisted quantum LDPC code
depends not only on order $v$ and block size $\mu$ but also on $2$-ranks concerning the parity-check matrix we use for decoding
because, as Theorem \ref{paraealdpc} states,
it is $\frac{v(v-1)}{\mu(\mu-1)}-2\operatorname{rank}(H)+\operatorname{rank}(HH^T)$ for a given incidence matrix $H$ of an $S(2,\mu,v)$.
The known results on the possible values of $2$-ranks of $S(2,\mu,v)$s were reviewed
in the context of entanglement-assisted quantum LDPC codes in \cite{Fujiwara:2010e,Fujiwara:2013b}.
For convenience, we summarize useful known results here.

The following are the explicit formulas of the $2$-ranks of highly even-free projective geometric $S(2,\mu,v)$s discussed in Section \ref{ldpcphasesec}.
\begin{theorem}[\cite{Hamada:1968}]\label{PGqeven}
Let $H$ be an incidence matrix of an $S(2,\mu,v)$ that forms the points and $1$-dimensional subspaces of $\textup{PG}(m,q)$ with $q$ even.
Define $t = \log_2q$.
The $2$-rank $\operatorname{rank}(H) = \varphi_{e}(m,q)$ of the incidence matrix $H$ is given by
\begin{align*}&\varphi_e(m,q) =\\
&\sum_{(s_{0}, s_{1}, \ldots, s_{t})}\prod_{j=0}^{t-1}\sum_{i=0}^{L(s_{j+1},s_{j})}l^i{m+1 \choose i}{m+2s_{j+1}-s_{j}-2i \choose m}
\end{align*}
where $l = -1$, the sum is taken over all ordered sets
$(s_{0}, s_{1}, \ldots, s_{t})$ with $s_{0} = s_{t}$, $s_{j} \in \mathbb{N}_0$ such that $0 \leq s_{j} \leq m-1$
and $0 \leq 2s_{j+1} - s_{j} \leq m+1$ for each $j = 0, \ldots, t-1$, and \[L(s_{j+1},s_{j}) = \left\lfloor\frac{2s_{j+1} - s_{j}}{2}\right\rfloor.\]
\end{theorem}
\begin{theorem}[\cite{Frumkin:1990a}]\label{rankPGqodd}
Let $H$ be an incidence matrix of an $S(2,\mu,v)$ that forms the points and $1$-dimensional subspaces of $\textup{PG}(m,q)$ with $q$ odd.
Then
\[\operatorname{rank}(H) = v-1 = \frac{q^{m+1}-q}{q-1}.\]
\end{theorem}

The $2$-rank for the case of a highly even-free Steiner $2$-design
forming the points and $1$-dimensional affine subspaces of $\textup{AG}(m,q)$ with $q$ even can be expressed by
$\varphi_e(m,q)$, that is, the $2$-rank of an incidence matrix of an $S(2,\mu,v)$ based on $\textup{PG}(m,q)$ with $q$ even.
\begin{theorem}[\cite{Hamada:1973}]\label{rankAGqeven}
Let $H$ be an incidence matrix of an $S(2,\mu,v)$ that forms the points and $1$-dimensional affine subspaces of $\textup{AG}(m,q)$ with $q$ even.
Then the $2$-rank of $H$ is given by
\[\operatorname{rank}(H) = \varphi_e(m,q) - \varphi_e(m-1,q).\]
\end{theorem}

If $q$ is odd, the $2$-rank for the case of $\textup{AG}(m,q)$ is full.
\begin{theorem}[\cite{Yakir:1993}]\label{rankAGqodd} Let $H$ be an incidence matrix of an $S(2,\mu,v)$ formed by
the points and $1$-dimensional affine subspaces of $\textup{AG}(m,q)$ with $q$ odd.
Then
\[\operatorname{rank}(H) = v = q^{m}.\]
\end{theorem}

If one wishes to employ an $S(2,\mu,v)$ that is not the points and $1$-dimensional subspaces of $\textup{PG}(m,q)$ or
the points and $1$-dimensional affine subspaces of $\textup{AG}(m,q)$,
it is necessary to know the $2$-rank of its incidence matrix to compute the dimension through Theorem \ref{paraealdpc}.
The following are two results on the $2$-ranks of $S(2,\mu,v)$s applicable to half of all general cases.
\begin{theorem}[\cite{Hamada:1973}]\label{rank1}
If $\frac{\mu(v-\mu)}{\mu-1}$ is odd, then any incidence matrix $H$ of an $S(2,\mu,v)$ is of full rank, that is, $\operatorname{rank}(H) = v$.
\end{theorem}
\begin{theorem}[\cite{Hamada:1973}]\label{rank2}
If $\mu$ is even and $\frac{v-\mu}{\mu-1}$ is odd, then for any incidence matrix $H$ of an $S(2,\mu,v)$,
$\operatorname{rank}(H) = v-1$.
\end{theorem}

When $\frac{v-\mu}{\mu-1}$ is even, the $2$-ranks of incidence matrices of $S(2,\mu,v)$s may take various values even if $v$ and $\mu$ are fixed.
In fact, they may vary if Steiner $2$-designs are not mutually isomorphic.
Hence, if $\frac{v-\mu}{\mu-1}$ is even, finer structural information than the order and block size is needed to calculate the dimension.
The most general bounds on the $2$-rank of an $S(2,\mu,v)$ read as follows.
\begin{theorem}[\cite{Hillebrandt:1992}]\label{bound:rkSteiner}
The $2$-rank $\operatorname{rank}(H)$ of an incidence matrix $H$ of an $S(2,\mu,v)$ satisfies inequalities
\[ \left\lceil \frac{1}{2}+\sqrt{\frac{1}{4}+\frac{(v-1)(v-\mu)}{\mu}}\right\rceil \leq \operatorname{rank}(H) \leq v.\]
\end{theorem}

The following is a very strong theorem for the case when the block size $\mu$ is $3$.
\begin{theorem}[\cite{Assmus-Jr.:1995}]\label{STSAssmus}
For any $v \equiv 3, 7 \pmod{12}$, where $v = 2^tu-1$ and $u$ is odd,
and any integer $i$ with $1 \leq i < t$,
there exists an $S(2,3,v)$ whose incidence matrix $H$ satisfies the condition that $\operatorname{rank}(H) = v-t+i$.
\end{theorem}

It is notable that the theorem above covers all orders $v$ to which Theorem \ref{rank1} is not applicable.
It is also worth noting that the machinery behind Theorem \ref{STSAssmus} can construct any $S(2,3,v)$ whose incidence matrix $H$
satisfies the condition that $\operatorname{rank}(H) \leq v-1$.
While the theorem does not treat the case $\operatorname{rank}(H) = v$,
the vast majority of $S(2,3,v)$s are actually of full rank.
The following theorem provides a simple way to find such Steiner $2$-designs.
\begin{theorem}[\cite{Doyen:1978}]\label{cyclicrank}
Let $H$ be an incidence matrix of an $S(2,3,v)$ with a transitive automorphism group.
Then $\operatorname{rank}(H) = v$ except when the $S(2,3,v)$ is the points and $1$-dimensional subspaces of $\textup{PG}(m,2)$.
\end{theorem}

For instance, it is known that for all $v \equiv 1, 3 \pmod{6}$ except for $9$ there exists an $S(2,3,v)$
in which the cyclic group of order $v$ acts regularly on the points \cite{Colbourn:1999}.
Such an $S(2,3,v)$ is called \textit{cyclic}.
By Theorems \ref{cyclicrank}, a cyclic $S(2,3,v)$ always gives an incidence matrix of full rank
except when it is the points and $1$-dimensional subspaces of a projective geometry over the binary field $\mathbb{F}_2$.
The $2$-rank of an incidence matrix of an $S(2,3,v)$ from $\textup{PG}(m,2)$ is known as well.
\begin{theorem}[\cite{Doyen:1978}]\label{cyclicexception}
Let $H$ be an incidence matrix of an $S(2,3,2^{m+1}-1)$.
Then $\operatorname{rank}(H) \geq 2^{m+1}-m-2$
with equality if and only if the $S(2,3,2^{m+1}-1)$ is the points and $1$-dimensional subspaces of $\textup{PG}(m,2)$.
\end{theorem}

To compute the dimensions of our entanglement-assisted quantum LDPC codes constructed through Theorem \ref{paraealdpc} with Steiner $2$-designs,
we also need to know the $2$-rank $\operatorname{rank}(HH^T)$ for a given incidence matrix $H$ of an $S(2,\mu,v)$.
An important fact to note is that this number $\operatorname{rank}(HH^T)$ is also exactly the number of perfectly noiseless qubits we need to provide.
The case when $\operatorname{rank}(HH^T) = 0$ reduces to the case of standard stabilizer codes, where $4$-cycles inevitably appear in the Tanner graph of $H$.
Since it is our aim to minimize the number of required ebits,
our primary focus is on those highly even-free $S(2,\mu,v)$s and similar promising combinatorial designs
whose incidence matrices $H$ satisfy the condition that $\operatorname{rank}(HH^T) = 1$.
It should be noted that this does not mean that we should dismiss entanglement-assisted quantum error-correcting codes requiring more than one ebit.
If the required number of ebits is reasonably small or if theoretically interesting phenomena can be found,
it is equality of interest to investigate the case when $\operatorname{rank}(HH^T) > 1$.
In this paper, however, we limit ourselves to the single ebit assistance, where combinatorial tools can be exploited effectively.

Recall that an $S(2,\mu,v)$ is a special $\textup{PBD}(v,K,1)$ with $K = \{\mu\}$.
The \textit{replication number} $r_x$ of a point $x \in V$ of a PBD $(V, {\mathcal B})$
is the number of occurrences of $x$ in the blocks of ${\mathcal B}$.
A PBD is \textit{odd-replicate} if for every $x \in V$ the replication number $r_x$ is odd.
If the replication number is even for every point, it is \textit{even-replicate}.
If $r_x = r_y$ for any two points $x$ and $y$,
we say that the PBD is \textit{equireplicate} (or \textit{regular}) and has replication number $r_x$.
Every $S(2,\mu,v)$ is equireplicate and has replication number $\frac{v-1}{\mu-1}$.
Note that while incidence matrices of regular PBDs result in parity-check matrices of \textit{right-regular} LDPC codes in the language of coding theory,
their column weights are not necessarily uniform, which means that they may not be \textit{left-regular}.
To avoid any confusion, we use the term equireplicate instead of regular when referring to combinatorial designs.

The following is our basic tool to identify combinatorial designs that require as few ebits as possible.
\begin{theorem}[\cite{Fujiwara:2013b}]\label{oddreplicate}
Let $H$ be a matrix over $\mathbb{F}_2$ in which every row and column is of weight greater than $1$.
$\operatorname{rank}(HH^T) = 1$ if and only if $H$ is an incidence matrix of an odd-replicate $\textup{PBD}(v,K,1)$
in which every point appears more than one block and no block is of size $1$.
\end{theorem}

Assuming that we exclude trifling examples such as LDPC codes of minimum distance $1$ or $2$,
the above theorem essentially says that the number of required ebits is minimized if and only if
we use an odd replicate PBDs of index $1$.
If we limit ourselves to $S(2,\mu,v)$s, this means that incidence matrices $H$ of those with $\frac{v-1}{\mu-1}$ odd
meet the condition that $\operatorname{rank}(HH^T)=1$.

While a significant portion of $S(2,\mu,v)$s including many highly even-free ones given in Section \ref{ldpcphasesec} are indeed odd-replicate,
not all Steiner $2$-designs are.
If one wishes to employ even-replicate $S(2,\mu,v)$s as well while not requiring many ebits,
a naive and straightforward way would be to join the identity matrix $I$ to an incidence matrix $H$
to form $H' = \left[\begin{array}{cc}I & H\end{array}\right]$ as we did for quantum error-correcting codes assisted by less noisy qubits.
If we reindex the rows and columns of the extended matrix $H'$ by blocks and points,
because the blocks of size $1$ corresponding to the columns of $I$ contain no pair of points,
$H'$ forms an incidence matrix of an odd-replicate $\textup{PBD}(v,K,1)$.
It is easy to verify that the number $\operatorname{rank}(H'{H'}^T)$ of required ebits becomes $1$.
The problem of this approach is that the minimum distance of the resulting LDPC code is always $\mu+1$
regardless of the even-freeness of the $S(2,\mu,v)$ defined by $H$.
Fortunately, because parity-check matrices do not need to be in standard form in entanglement-assistance,
there is a simple way around this problem so that one may exploit the promising structure of an incidence matrix of an $S(2,\mu,v)$ even if it is even-replicate.
\begin{theorem}\label{addR}
Let $H$ be an incidence matrix of an even-replicate $S(2,\mu,v)$ and $\mu \geq 2$.
Take the $v \times v$ identity matrix $I$, the $v$-dimensional all-one vector $J_{1,v} = (1,\dots,1)$,
and the $\frac{v(v-1)}{\mu(\mu-1)}$-dimensional all-zero vector $\boldsymbol{0}_{1,\frac{v(v-1)}{\mu(\mu-1)}}$.
Define a $(v+1)\times\left(\frac{v(v-1)}{\mu(\mu-1)}+v\right)$ matrix $H'$ as
\begin{align*}
H' &= \left[\begin{array}{cc}I & H\\
J_{1,v} & \boldsymbol{0}_{1,\frac{v(v-1)}{\mu(\mu-1)}}\end{array}\right]\\
&= \left[\begin{array}{cc}I & H\\
\text{\small{1 \dots 1}} & \text{{\small 0 \dots 0}}\end{array}\right].
\end{align*}
$H'$ is an incidence matrix of a $\textup{PBD}(v+1,K,1)$ such that $\operatorname{rank}(H'{H'}^T) = 1$.
In particular, if the original $S(2,\mu,v)$, $(V,\mathcal{B})$, is $r$-even-free
and satisfies the property that for any nonempty configuration $\mathcal{C} \subseteq \mathcal{B}$ with $\vert\mathcal{C}\vert \leq r$
and $\operatorname{odd}(\mathcal{C})$ even
\begin{align}\label{extensioncondition}
\vert \mathcal{C} \vert + \operatorname{odd}(\mathcal{C}) \geq r+1,
\end{align}
where $\operatorname{odd}(\mathcal{C})$ is the number of points $v \in V$ such that $v$ is contained in an exactly odd number of blocks in $\mathcal{C}$,
then the $\textup{PBD}(v+1,K,1)$ is $r$-even-free.
\end{theorem}
\begin{IEEEproof}
Take an element $\infty \not\in V$ and define a finite set $V' = V \cup \{\infty\}$ of size $\vert V'\vert = v+1$.
Index the rows of $H'$ by the elements of $V'$ such that
the additional element $\infty$ is associated with the row  $(J_{1,v}, \boldsymbol{0}_{1,\frac{v(v-1)}{\mu(\mu-1)}})$
and such that the other $v$ rows are associated the same way as in the incidence matrix $H$.
For every unordered pair $\{\infty, v\}$ with $v \in V$,
there exists an unique column of $H'$ in which the rows corresponding to $\infty$ and $v$ both contain $1$.
Because $H$ is an incidence matrix of a Steiner $2$-design,
for every unordered pair $\{v, w\}$ such that $v, w \in V$ there exists exactly one column in which
the rows indexed by $v$ and $w$ simultaneously contain $1$.
Hence, the extended matrix $H'$ is an incidence matrix of a $\textup{PBD}(v+1,K,1)$ with $V'$ as its point set,
where the block set $\mathcal{B}'$ consists of $v$ blocks of size $2$ and $\frac{v(v-1)}{\mu(\mu-1)}$ blocks of size $\mu$.
It suffices to show that the resulting PBD does not contain any even configurations of size $r$ or smaller
if the original $S(2,\mu,v)$ is $r$-even-free and
if every nonempty configuration $\mathcal{C} \subseteq \mathcal{B}$ of size $r$ or smaller
such that $\operatorname{odd}(\mathcal{C})$ is even satisfies the inequality
$\vert \mathcal{C} \vert + \operatorname{odd}(\mathcal{C}) \geq r+1$.
Suppose to the contrary that the PBD $(V',\mathcal{B}')$ contains an even configuration $\mathcal{D}$ of size smaller than or equal to $r$.
Define $\mathcal{D}_H = \{B \in \mathcal{D} \mid B \in \mathcal{B}\}$ to be the set of blocks in $\mathcal{D}$ that are also contained in $\mathcal{B}$.
If $\operatorname{odd}(\mathcal{D}_H)$ is odd, the number of blocks in $\mathcal{D}$ that contain $\infty$ is odd,
contradicting the assumption that $\mathcal{D}$ is an even configuration.
If $\operatorname{odd}(\mathcal{D}_H)$ is even, by assumption,
$\vert \mathcal{D}_H \vert + \operatorname{odd}(\mathcal{D}_H) \geq r+1$.
However, because $\operatorname{odd}(\mathcal{D}_H) = \vert\mathcal{D}\setminus\mathcal{D}_H\vert$,
we have
\begin{align*}
\vert \mathcal{D}_H \vert + \operatorname{odd}(\mathcal{D}_H) &= \vert \mathcal{D}_H \vert + \vert\mathcal{D}\setminus\mathcal{D}_H\vert\\
&= \vert\mathcal{D}\vert\\
&\leq r,
\end{align*}
a contradiction. The proof is complete.
\end{IEEEproof}

Note that the resulting parity-check matrix in the above theorem is larger and has more nonzero entries than the original.
This may slightly increase the decoding complexity, although the density will still be in the decodable range unless
the original parity-check matrix is already barely decodable by an iterative decoding algorithm.

Now we illustrate how to apply various theorems presented here
and demonstrate how to effectively take advantage of the theorem above through an example case.
We first construct through results given in this paper
a known class of good entanglement-assisted quantum LDPC codes which originally appeared in \cite{Fujiwara:2010e}.
Then we show how Theorem \ref{addR} extends the class.

By Theorem \ref{affineerasure}, the points and $1$-dimensional affine subspaces of $\textup{AG}(m,q)$ with $q$ odd
form the points and blocks of a $(2q-1)$-even-free $S(2,q,q^m)$, which achieves the upper bound on the even-freeness given in Theorem \ref{boundabelian}.
Because the replication number of an $S(2,q,q^m)$ is
\[\frac{q^m-1}{q-1} = \sum_{i=0}^{m-1}q^i,\]
it is odd-replicate if $m$ is odd, ensuring that the number of required ebits is $1$ by Theorem \ref{oddreplicate}.
Hence, considering also its girth and extremely high rate as an $S(2,\mu,v)$, which was discussed in Section \ref{ldpcphasesec},
it would not be too optimistic to expect that
an incidence matrix of the $S(2,q,q^m)$ would work well as a parity-check matrix for an entanglement-assisted quantum LDPC code.
Applying the rank formula given in Theorem \ref{rankAGqodd} to Theorem \ref{paraealdpc},
the corresponding entanglement-assisted quantum LDPC code is of length $q^{m-1}\frac{q^m-1}{q-1}$ and dimension $q^{m-1}\frac{q^m-1}{q-1}-2q^m$,
namely a $[[q^{m-1}\frac{q^m-1}{q-1}, q^{m-1}\frac{q^m-1}{q-1}-2q^m]]$ quantum error-correcting code
that allows for quantum error correction through the LDPC code of length $q^{m-1}\frac{q^m-1}{q-1}$, dimension $q^{m-1}\frac{q^m-1}{q-1}-q^m$,
girth $6$, and minimum distance $2q$ defined by an incidence matrix of the $S(2,q,q^m)$.

The above class of codes based on $\textup{AG}(m,q)$ is indeed known to perform in simulations quite similarly
to finite geometry LDPC codes proposed in \cite{Hsieh:2011} (see \cite{Clark:2011}).
As we have just seen, however, if a large number of ebits are prohibited,
this straightforward method only admits $\textup{AG}(m,q)$ with $m$ odd when $q$ is also odd.
Theorem \ref{addR} provides a means to exploit the even-replicate $S(2,q,q^m)$ based on $\textup{AG}(m,q)$ with $m$ even.
In fact, Theorem \ref{AGoddpointtheorem} assures that for any $m$ the $(2q-1)$-even-free $S(2,q,q^m)$ from $\textup{AG}(m,q)$ with $q$ odd
satisfies a stronger condition than Inequality (\ref{extensioncondition}) in Theorem \ref{addR}.
Thus, its parity-check matrix $H$ can be extended to a $(q^m+1) \times (q^{m-1}\frac{q^m-1}{q-1} + q^m)$ matrix $H'$
which forms an incidence matrix of a $(2q-1)$-even-free $\textup{PBD}(q^m+1,K,1)$.
Because $\operatorname{rank}(H'{H'}^T) = 1$, the entanglement-assisted quantum LDPC code based on this new $\textup{PBD}$ requires only $1$ ebit,
as opposed to $q^m-1$ ebits in the case of the straightforward use of $\textup{AG}(m,q)$ with $m$ even and $q$ odd
(see \cite{Fujiwara:2010e} for the formula for $\operatorname{rank}(HH^T)$ of Steiner $2$-designs).
Note that the $2$-rank of $H'$, which is required to know to compute the dimension of our code based on the $\textup{PBD}$,
can be easily obtained. Indeed, it is simply of full rank, that is, $\operatorname{rank}(H') = q^m+1$.
This is because the first $q^m$ rows $\left[\begin{array}{cc}I & H\end{array}\right]$ are linearly independent due to the identity matrix $I$
and also because no linear combination of these rows coincides with the bottom row
\[\left[\begin{array}{cc} J_{1,q^m} & \boldsymbol{0}_{1,\frac{q^{m-1}(q^m-1)}{q-1}}\end{array}\right]\]
due to the fact that $q$ is odd.
In fact, all of the first $q^m$ rows must be added together to obtain $J_{1,q^m}$ on the left-hand side
while adding up all of them results in the all-one vector on the right-hand side instead of the required zero vector.
The fundamental parameters of the new entanglement-assisted quantum LDPC code can be summarized as follows.
\begin{corollary}\label{AGparacoro}
For any even integer $m \geq 2$ and odd prime power $q$ there exists an entanglement-assisted quantum error-correcting code
of length $q^{m-1}\frac{q^m-1}{q-1} + q^m$ and dimension $q^{m-1}\frac{q^m-1}{q-1}-q^m-1$
that requires exactly $1$ ebit and can be decoded by the LDPC code of length $q^{m-1}\frac{q^m-1}{q-1} + q^m$,
dimension $q^{m-1}\frac{q^m-1}{q-1}-1$, girth $6$, and minimum distance $2q$.
\end{corollary}

It is notable that quantum error-correcting codes constructed through Theorem \ref{addR} are generally expected to have higher dimensions
than the original codes used as ingredients because of the extra columns.
In the next section we will demonstrate through simulations that this type of quantum LDPC code performs well as expected.

\section{Simulation Results}\label{sim}
In this section we report simulation results on the performance of our quantum LDPC codes.
As noted earlier at the beginning of Section \ref{LDPC}, we try to be conservative and assume that
no exploitable information is available to the receiver regarding possible correlations between bit errors and phase errors.
Thus, decoding is done separately for bit errors and phase errors through the sum-product algorithm
over two independent binary symmetric channels,
where one channel introduces the operator $X$ independently on each physical qubit with probability $p_x$
and the other causes the operator $Z$ to act independently on each physical qubit with probability $p_z$.
Error correction succeeds if the decoder correctly identifies all qubits on which the $X$ operator acted
through the sum-product algorithm over one binary symmetric channel
and also properly locates all qubits suffering from the $Z$ operator the same way over the other binary symmetric channel.

As in \cite{MacKay:2004}, we report the block error rate (BLER) $b_p$
of our LDPC codes over the binary symmetric channel with crossover probability $p$.
Thus, for instance, if one would like a conservative estimate on the performance of the corresponding quantum LDPC codes
over the depolarizing channel with equal error probability $\frac{p}{2}$ for each of the three types of quantum error,
the estimated BLER over the quantum channel is $1-(1-b_p)^2 \approx 2b_p$.
Over a more general Pauli channel with a small error probability $p_y$ for the Pauli operator $Y$,
the same calculation gives a reasonable estimate on the performance of our quantum LDPC codes.

We compare our quantum LDPC codes with \textit{hypothetical} ones that would be available
through the CSS construction if there were no constraint on the structure of a parity-check matrix.
More specifically, we compete with the ideal situation where any parity-check matrix $H$ of an LDPC code,
whether it is in standard form or not, can be used
to form a quantum LDPC code regardless of the value of $\operatorname{rank}(HH^T)$.
Hence, any parity-check matrix of any linear $[n,k,d]$ code gives rise to a hypothetical $[[n,2k-n]]$ quantum error-correcting code.

For parity-check matrices of hypothetical codes,
we chose good incidence matrices of combinatorial designs found in the coding theory literature
and those obtained through the progressive edge-growth (PEG) algorithm,
which is among the most successful known methods for designing LDPC codes of relatively short length in the classical domain \cite{Hu:2005}.
Comparison against the structured LDPC codes from promising combinatorial designs
makes it easy to see how much the matrix extension processes in Theorems \ref{quantumS} and \ref{addR} affect
the performance in our context while the PEG algorithm provides good hypothetical codes for general performance comparison purposes.

\begin{figure}[t!]
	\includegraphics[scale = 0.75, trim=125 225 150 238, clip]{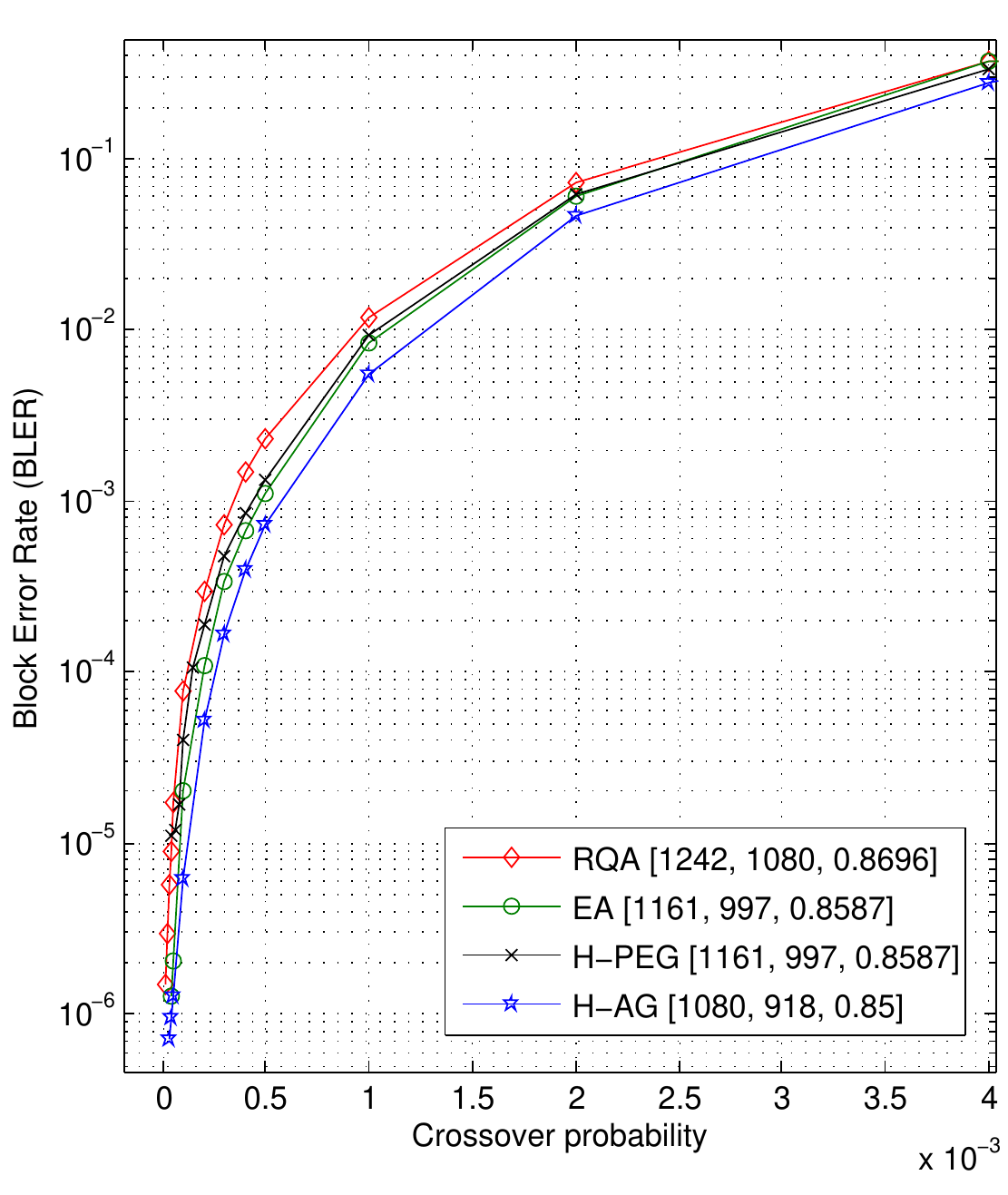}
	\caption{Block error rates of quantum LDPC codes obtained from $\textup{AG}(4,3)$. $\vert$ RQA and EA refer to assistance by less noisy qubits and
	ebits respectively. H-PEG and H-AG stand for hypothetical CSS codes based on classical LDPC codes
	generated by the PEG algorithm and affine geometry $\textup{AG}(4,3)$ respectively. The parameters are shown in square brackets
	in order of $[\text{length}, \text{dimension}, \text{rate}]$.}
	\label{fig1}
\end{figure}
\begin{figure}[t!]
	\includegraphics[scale = 0.75, trim=125 225 150 238, clip]{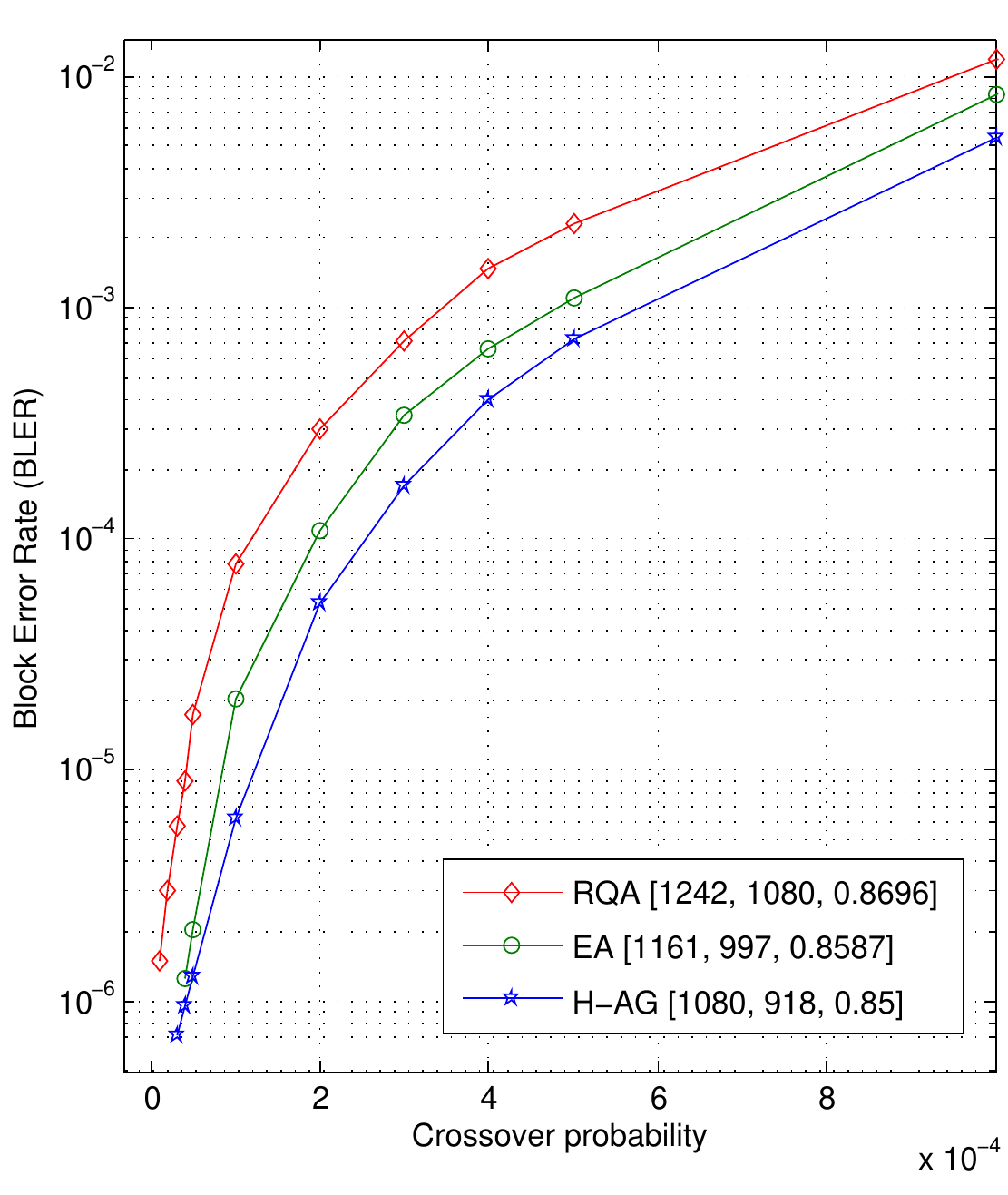}
	\caption{Block error rates of quantum LDPC codes obtained from $\textup{AG}(4,3)$ at lower crossover probabilities. $\vert$
	RQA and EA refer to assistance by less noisy qubits and ebits respectively.
	H-AG stands for the hypothetical CSS code based on $\textup{AG}(4,3)$ that would be available if there were no structural constraint.
	The parameters are shown in square brackets in order of $[\text{length}, \text{dimension}, \text{rate}]$.}
	\label{fig2}
\end{figure}

Fig.\ \ref{fig1} shows the block error rates of our quantum LDPC codes obtained from $\textup{AG}(4,3)$ through Theorems \ref{quantumS} and \ref{addR},
the hypothetical CSS code based on $\textup{AG}(4,3)$, and another hypothetical one generated by the PEG algorithm.
The parameters of these codes are summarized in Table \ref{tb1}.
\begin{table*}
\renewcommand{\arraystretch}{1.3}
\caption{Parameters of quantum LDPC codes}
\label{tb1}
\centering
\begin{tabular}{ccccccccc}
\hline\hline
{\bfseries Type}\rlap{\textsuperscript{a}}&\bfseries Length &{\bfseries Dimension}\rlap{\textsuperscript{b}}
&{\bfseries Reliable Qubit}&{\bfseries Ebit}\rlap{\textsuperscript{c}}&{\bfseries Rate}&{\bfseries Mean Column/Row Weight}&{\bfseries Max Column/Row Weight}
&{\bfseries Distance}\rlap{\textsuperscript{d}}\\
\hline
RQA & $1242$ & $1080$ & $162$ & $0$ & $0.8696$ & $2.86/41$ & $3/41$ & $4$\\
EA & $1161$ & $997$ & $0$ & $1$ & $0.8587$ & $2.93/41.48$& $3/81$& $6$\\
H-PEG & $1161$ & $997$ & 0 & (81) &$0.8587$ & $3/42.47$ & $3/43$& $4$\\
H-AG & $1080$ & $918$ & 0 & (80) & $0.85$ & $3/40$ & $3/40$ & $6$\\
 \hline
 \hline
\multicolumn{9}{l}{\scriptsize\textsuperscript{a}
This column shows whether it is assisted by qubits with possible phase errors (RQA),
assisted by ebits with no errors (EA), a hypothetical CSS code generated by the}\vspace{-1.1mm}\\
\multicolumn{9}{l}{\scriptsize\phantom{\textsuperscript{a}}
PEG algorithm (H-PEG), or a hypothetical one generated by $\textup{AG}(4,3)$ (H-AG).}\vspace{-1.1mm}\\
\multicolumn{9}{l}{\scriptsize\textsuperscript{b} The EA code is in catalytic mode for fair comparison. For details, see \cite{Brun:2006d}.}\vspace{-1.1mm}\\
\multicolumn{9}{l}{\scriptsize\textsuperscript{c} Parentheses indicate the number of ebits required if Theorem \ref{eaprop} were applied.}\vspace{-1.1mm}\\
\multicolumn{9}{l}{\scriptsize\textsuperscript{d} Degeneracy and harmless nontrivial operators are taken into account,
so that this column shows the true distance of each quantum error-correcting code.}\vspace{4.4mm}
\end{tabular}
\end{table*}
A close-up of the three based on $\textup{AG}(4,3)$ at a small crossover probability region is given in Fig.\ \ref{fig2}.
As shown in these figures and the table,
our assisted codes exhibit block error rates comparable to those of the hypothetical ones
while significantly reducing the difficulty in implementation and slightly improving information rates.

\begin{figure}[t!]
	\includegraphics[scale = 0.75, trim=125 225 149 238, clip]{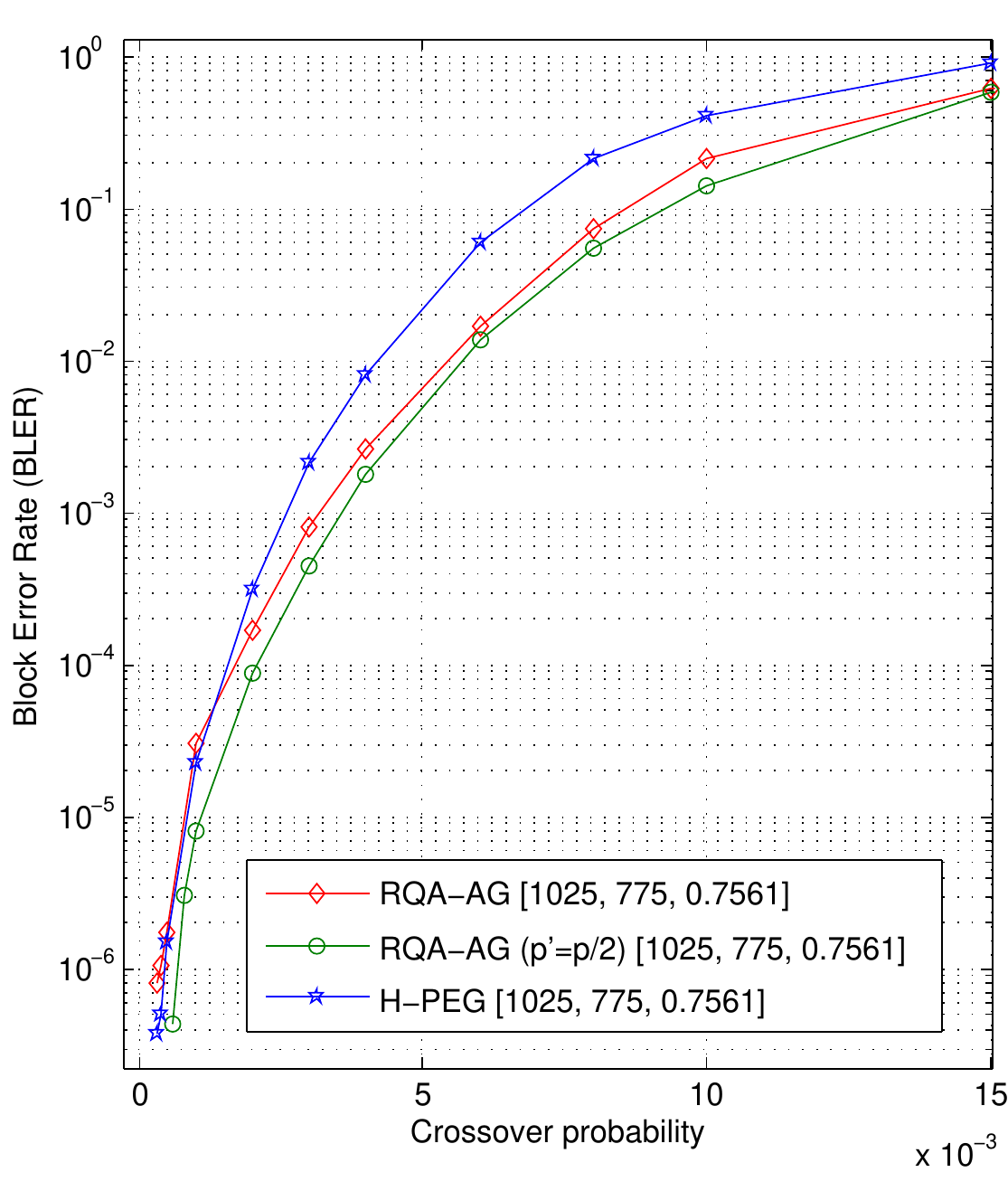}
	\caption{Comparision between quantum LDPC codes from $\textup{AG}(3,5)$ assisted by less noisy qubits
	and a hypothetical one generated by the PEG algorithm. $\vert$
	RQA-AG refers to the quantum LDPC code based on $\textup{AG}(3,5)$ assisted by less noisy qubits.
	H-PEG refers to a hypothetical one that would be available if the CSS construction did not impose orthogonality on a parity-check matrix.
	Also plotted are block error rates for when each less noisy qubit of RQA-AG experiences a phase error independently with probability $p' = \frac{p}{2}$,
	where a phase error occurs on each nosy qubit independently with crossover probability $p$.
	The parameters are shown in square brackets in order of $[\text{length}, \text{dimension}, \text{rate}]$.}
	\label{fig3}
\end{figure}
Fig.\ \ref{fig3} compares our quantum LDPC code from $\textup{AG}(3,5)$ assisted by less noisy qubits
with a hypothetical one constructed by the PEG algorithm.
These are both $[[1025,775]]$ quantum error-correcting codes of rate approximately $0.75$.
The former requires $250$ of all $1025$ qubits to be free from bit errors.
The latter would need, if maximally entangled pairs were to be used as ebits for quantum error correction,
$122$ qubits on the sender side which were maximally entangled to another set of $122$ perfectly noiseless qubits on the receiver side.
We also plotted block error rates of our code when the $250$ less noisy auxiliary qubits experience phase errors less frequently than the rest.
This additional assumption can be reasonable because the auxiliary qubits are supposed to be engineered more reliably and protected carefully.
As an example, we examined the case
when the phase error probability of each auxiliary qubit is a half of that of each noisy one.

\begin{figure}[t!]
	\includegraphics[scale = 0.69, trim=110 193 133 206, clip]{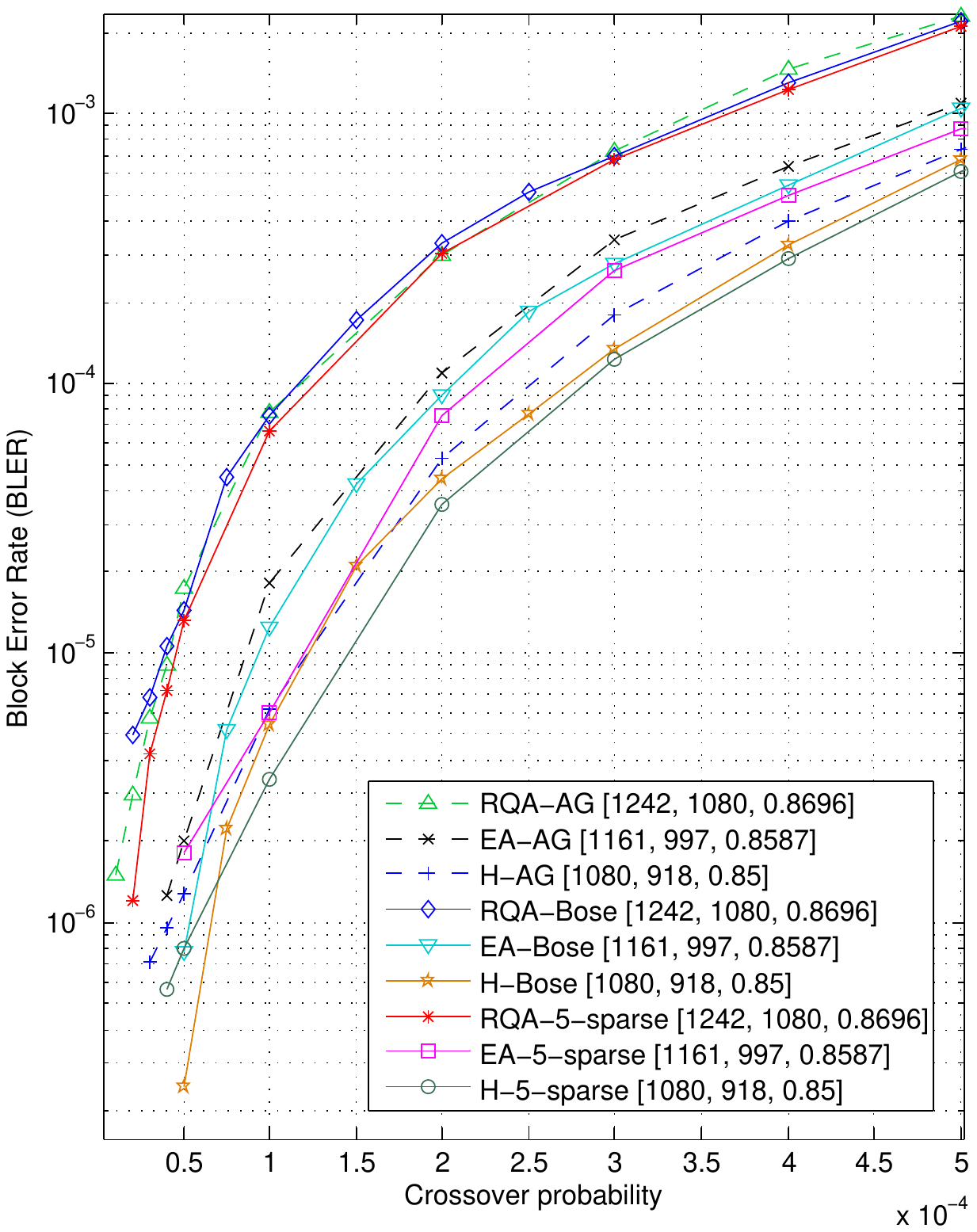}
	\caption{Block error rates of quantum LDPC codes from different Steiner $2$-designs of the same parameters. $\vert$
	RQA and EA refer to assistance by less noisy qubits and ebits respectively.
	H indicates that a hypothetical situation is assumed where no constraint is imposed on a parity-check matrix by the CSS construction.
	Codes from $\textup{AG}(4,3)$, a Bose-type Kirkman triple system, and a $5$-sparse Steiner triple system are shown.
	All combinatorial designs form $S(2,3,81)$s.
	The parameters of the corresponding quantum LDPC codes are shown in square brackets in order of $[\text{length}, \text{dimension}, \text{rate}]$.}
	\label{fig4}
\end{figure}
Different Steiner $2$-designs of the same order and same block size are compared in Fig.\ \ref{fig4}.
Simulation results of LDPC codes from $S(2,3,81)$s that form \textit{Kirkman triple systems} constructed through the \textit{Bose construction} \cite{Colbourn:1999}
and $5$\textit{-sparse Steiner triple systems} given in \cite{Colbourn:1994d} are presented along with the results of those from $\textup{AG}(4,3)$.
The $S(2,3,v)$s from the Bose construction including Kirkman triple systems were studied
for use as LDPC codes over additive white Gaussian noise channels in \cite{Vasic:2004,Johnson:2003}.
Incidence matrices of $5$-sparse Steiner triple systems are known to avoid small configurations
harmful to iterative decoding over a binary erasure channel \cite{Colbourn:2009}.
While they are not designed specifically for a binary symmetric channel,
LDPC codes that are good for these major channels typically perform fairly well,
especially if common harmful structures such as short cycles are avoided.
For an analysis of the effects of small configurations in $S(2,3,v)$s on iterative decoding,
the interested reader is referred to \cite{Laendner:2010}.
As expected, our simulation results are, while not identical, overall similar across different $S(2,3,v)$s for each type of quantum LDPC code.

\section{Concluding Remarks}\label{conc}
We explored the use of quantum error correction assisted by reliable qubits in the context of iterative decoding and
demonstrated how one may exploit combinatorics and known designing methods for structured LDPC codes.
The range of exploitable classical error-correcting codes for quantum error correction is extended by taking advantage of
the fact that some kind of quantum noise is easier to suppress on hardware by physical means.
A simple method for creating parity-check matrices of quantum LDPC codes assisted by only one ebit is also given.
These codes are shown to have error correction performance comparable to what would be achievable through the same classical ingredients
if the CSS construction did not impose any constraints on parity-check matrices.

It should be noted, however, that our results do not imply that
the approach presented in this paper removed all difficulties in designing quantum LDPC codes
or that the combination of the sum-product algorithm and suitable parity-check matrices is always superior to other coding methods.
Rather, Theorem \ref{lessnoisylemma} and the idea behind Theorem \ref{addR} should be understood as
useful tools to circumvent hurdles in designing a variety of quantum error-correcting codes.

To illustrate one limitation in employing LDPC codes for quantum error correction,
consider the CSS codes constructed from dual-containing Bose-Chaudhuri-Hocquenghem (BCH) codes
(see \cite{Aly:2007} for the definition of dual-containing BCH codes and their use in quantum error correction).
As mentioned in \cite{MacKay:2004}, quantum BCH codes outperform all known quantum LDPC codes at rates above $0.8$
if the code length is allowed to be several thousand.
As far as the authors are aware, the situation does not change if we include all known entanglement-assisted quantum LDPC codes
that only require a reasonably small number of ebits.
This is partly because it is already not easy to design classical LDPC codes of very high rate that surpass the dual-containing BCH codes
in terms of block error rate in the finite length regime even if there is no structural constraint.
Hence, if we let less noisy qubits assist quantum error correction,
it would still be a challenging task to find a parity-check matrix in standard form
that outperforms a dual-containing BCH code of very high rate.

High performance BCH codes are particularly appealing
because they also have efficient decoding methods due to their cyclic property (see \cite{Grassl:2000} for decoding in the quantum case).
Classical codes with the same cyclic property are called \textit{cyclic codes}.
They are among the most widely used error-correcting codes in classical information transmission and computation.
Hence, it would be natural to ask if we can import cyclic codes including BCH codes as we did LDPC codes.

Fortunately, the answer is yes.
It is easy to see that Theorem \ref{lessnoisylemma} we proved in Section \ref{lessnoisy} is general enough to import BCH codes and other cyclic codes
with their efficient decoding methods and give quantum error-correcting codes with higher rates than
the straightforward quantum analogues of cyclic codes through the CSS construction.
In fact, as shown in the proof of the theorem, quantum error correction with less noisy qubits extracts the information about quantum noise
in the form of a syndrome and exploits it in the same way as in the standard decoding for a linear $[n,k,d]$ code.
Hence, in principle, any standard technique that infers errors from syndromes is directly exploitable in the quantum setting
as long as the number $2(n-k)$ of less noisy auxiliary qubits falls within the acceptable range.
In our case, because the linear codes in question are of classical information rate $\frac{k}{n} > 0.9$, only a small fraction of qubits need to be free from bit errors.
Thus, these BCH codes and other high-rate cyclic codes are ideal classical codes for quantum error correction via less noisy qubits.

The slight improvement in quantum information rate with this approach comes from the fact that
the standard CSS construction only generates an $[n,2k-n]$ quantum error-correcting code
as it is a special case of Theorem \ref{eaprop} when $\operatorname{rank}(HH^T)=0$.
As shown in Theorem \ref{lessnoisylemma}, assistance from less noisy qubits results in a $[[2n-k,k]]$ quantum error-correcting code.
Hence, we always have a slight gain
\begin{align*}
\frac{k}{2n-k}-\frac{2k-n}{n} &= \frac{2(n-k)^2}{n(2n-k)}\\
&> 0
\end{align*}
in information rate in the quantum domain.
It should be noted, however, that the higher rate comes at the expense of relative distance because
the number of qubits to be protected also slightly increases.

Finally, we point out two important questions we did not address in this paper.
One question we did not consider is how many auxiliary qubits should be allowed.
In the case of entanglement assistance, we limited ourselves to the extreme case where only one ebit is allowed.
While it is certainly better not to use more ebits in terms of feasibility of implementation,
as far as the authors are aware, there is no evidence that using exactly one ebit is significantly better than the best possible standard stabilizer codes or
entanglement-assisted ones that require a few ebits in terms of error correction performance in the finite length regime.
In the case of less noisy qubits, we allowed more auxiliary qubits.
Less noisy qubits would be easier to realize than completely noiseless ones.
However, it is not clear how many would be acceptable and whether it is always worth it to encode quantum information using Theorem \ref{binaryonlyZ}.
For instance, assume the extreme case where less noisy qubits are as easy to realize as those that may suffer both $X$ errors and $Z$ errors.
If this were the case, it would make more sense to only use less noisy qubits and encode quantum information by a code optimized for the phase damping channel.
While it is unlikely for such an extreme assumption to become realistic in the near future,
it is both natural and important to consider the break-even point where other coding schemes start to make more sense.

The other important aspect of quantum error correction we did not consider is the possible effects of degeneracy.
As is well-understood in quantum information theory, a nontrivial operator may happen to stabilize a given quantum state.
In the language of the stabilizer formalism, this is to say that a pair of operators are indeed indistinguishable from each other
if one is different from the other by an element of the stabilizer of encoded quantum information.
This implies that, for example, what may look a nontrivial error at first glance may turn out to have no effect on the encoded quantum information.
Thus, it is of importance to ask whether the tensor product of Pauli operators that corresponds to
a codeword of small weight in an underlying classical LDPC code in fact acts nontrivially on encoded quantum states.
If some turn out to be indistinguishable from the tensor product of the trivial operator $I$,
the weight of smallest uncorrectable operators of our quantum LDPC codes may be larger than
the minimum distances of the underlying classical LDPC codes.

More formally, the \textit{distance} of a quantum error-correcting code of length $n$
is the smallest weight of an undetectable nontrivial element of the Pauli group over $n$ qubits \cite{Nielsen:2000}.
A quantum error-correcting code of distance $d$ is called \textit{degenerate}
if at least one nontrivial element of weight smaller than $d$ in the Pauli group acts trivially on the encoded qubits due to degeneracy,
otherwise \textit{non-degenerate}.
For an in-depth treatment of the mechanism of possible degeneracy in the context of entanglement assistance, we refer the reader to \cite{Brun:2006d}.
In the remainder of our discussion on degeneracy, we assume that the reader is familiar with the entanglement-assisted stabilizer formalism as presented in the article.

Regarding degeneracy of our quantum LDPC codes, we conjecture that all codewords of sufficiently small weights in the classical LDPC codes we employed
indeed correspond to undetectable errors.
Our belief partly comes from the observation that it would be unlikely for a small weight codeword of an extremely high-rate LDPC code to be contained in its dual code,
which is necessarily of tiny dimension.
While we could not prove a general statement that would universally apply to all quantum LDPC codes we considered,
the following theorem confirms our intuition for the case when Theorem \ref{addR} is used to improve
the minimum distance of the LDPC code based on an even-replicate Steiner $2$-design of block size $3$.
\begin{theorem}\label{degeneracy}
Let $H$ be an incidence matrix of the even-replicate $S(2,3,v)$ such that the linear code that admits $H$ as its parity-check matrix is of minimum distance $d$.
Take the $v \times v$ identity matrix $I$, the $v$-dimensional all-one vector $J_{1,v} = (1,\dots,1)$,
and the $\frac{v(v-1)}{6}$-dimensional all-zero vector $\boldsymbol{0}_{1,\frac{v(v-1)}{6}}$.
Define a $(v+1)\times\left(\frac{v(v-1)}{6}+v\right)$ matrix $H'$ as
\begin{align*}
H' = \left[\begin{array}{cc}I & H\\
J_{1,v} & \boldsymbol{0}_{1,\frac{v(v-1)}{6}}\end{array}\right].
\end{align*}
The entanglement-assisted quantum error-correcting code $\mathcal{C}$ formed by the quantum parity-check matrix
\begin{align*}
\left[\begin{array}{c}\omega H'\\
\bar{\omega}H' \end{array}\right]
\end{align*}
is of distance at most $d$.
In particular, if the distance of $\mathcal{C}$ achieves the upper bound $d$ as in Theorem \ref{addR},
it is non-degenerate except possibly when $v = 21, 33, 45$.
\end{theorem}

To prove the above theorem, we use properties of the linear codes generated by the rows of incidence matrices.
Let $H$ be an incidence matrix of an $S(2,k,v)$.
The \textit{point code} of the $S(2,k,v)$ is the linear subspace over $\mathbb{F}_2$ spanned by the rows of $H$.
From the viewpoint of coding theory, it is simply the \textit{dual code} of the classical LDPC code whose parity-check matrix is $H$.
\begin{theorem}[\cite{Colbourn:2001}]\label{pointcodeColbourn}
The point code of an $S(2,3,v)$ contains a codeword of weight $w = \frac{v-1}{2} - \epsilon$ for $\epsilon > 0$
as a linear combination of $s$ rows for positive $s$ if and only if
one of the following holds.
\begin{enumerate}
\item $s = \frac{v+1}{2}$, $w \equiv 0 \pmod{4}$, and
\[w \geq \begin{cases}
0 & \text{if} \ \frac{v-1}{2} \equiv 1, 3 \pmod{6},\\
\frac{v-1}{3} & \text{if} \ \frac{v-1}{2} \equiv 0 \pmod{6},\\
\frac{v}{3}+1 & \text{if} \ \frac{v-1}{2} \equiv 4 \pmod{6}.
\end{cases}\]
\item $s = \frac{v+3}{2}$, $\frac{v-1}{2} \equiv 0, 4 \pmod{6}$, $w \equiv s \pmod{4}$, and
\[w \geq \begin{cases}
\frac{v+3}{6} & \text{if} \ \frac{v-1}{2} \equiv 4 \pmod{6},\\
\frac{v+35}{6} & \text{if} \ \frac{v-1}{2} \equiv 0 \pmod{6}.
\end{cases}\]
\item $s = \frac{v+5}{2}$, $\frac{v-1}{2} \equiv 1, 3 \pmod{6}$, $w \equiv 0 \pmod{4}$, and
\[w \geq \begin{cases}
\frac{v+5}{3} & \text{if} \ \frac{v-1}{2} \equiv 3 \pmod{6},\\
\frac{v+21}{3} & \text{if} \ \frac{v-1}{2} \equiv 1 \pmod{6}.
\end{cases}\]
\end{enumerate}
\end{theorem}
\begin{theorem}[\cite{Baartmans:1996}]\label{pointcodeTonchev}
A codeword of the point code of an $S(2,3,v)$ whose incidence matrix is $H$ is of weight $\frac{v-1}{2}$ only if it is either
\begin{enumerate}
\item a row of $H$,
\item a sum of $\frac{v-1}{2}$ rows of $H$ in which no block contains three of the corresponding $\frac{v-1}{2}$ points, or
\item a sum of $\frac{v-1}{2} + i$ rows of $H$, where $i = 1$ if $\frac{v-1}{2} \equiv 0 \pmod{4}$,
$i = 2$ if $\frac{v-1}{2} \equiv 1 \pmod{2}$, and $i = 3$ if $\frac{v-1}{2} \equiv 2 \pmod{4}$.
\end{enumerate}
\end{theorem}

\begin{IEEEproof}[Proof of Theorem \ref{degeneracy}]
It suffices to prove that the elements of the stabilizer of the entanglement-assisted quantum error-correcting code $\mathcal{C}$
that act trivially on the noiseless qubit do not include the tensor products of Pauli operators of weight $d$
whose noisy parts correspond to the minimum weight nonzero codewords of the linear code underlying $\mathcal{C}$.
Because the inner product of any pair of rows of $H'$ is $1$ and $\operatorname{rank}\big(H'{H'}^T\big) = 1$,
the generators that globally commute with each other and act trivially on the noiseless qubit forms the group
$\left\langle I\vert X^{\boldsymbol{r}}, I\vert Z^{\boldsymbol{r}} \ \middle\vert\ \boldsymbol{r}\in R_e\big(H'\big)\right\rangle$,
where $R_e\big(H'\big)$ is the set of linear combinations of even numbers of rows of $H'$
and the identity operator $I$ on the left to the vertical line represents the trivial action on the noseless qubit.
We show that the minimum weight nonzero codewords of the linear code underlying $\mathcal{C}$ are not in $R_e\big(H'\big)$.

The replication number of an $S(2,3,v)$ is $\frac{v-1}{2}$.
Hence, by Theorem \ref{pointcodeColbourn} and the assumption that the Steiner $2$-design is even-replicate,
the minimum distance of the point code of the $S(2,3,v)$ is bounded below by $\frac{v+3}{6}$.
Thus, the minimum distance of the linear code $\mathcal{L}$ spanned by the row of $H'$ is at least $\frac{v+3}{6}$ as well.
Note that $R_e\big(H'\big)$ is contained in $\mathcal{L}$,
which implies that $R_e\big(H'\big)$ does not contain nonzero vectors of weight less than $\frac{v+3}{6}$ either.
However, by Theorem \ref{th:paraSTSEAQECC}, the minimum distance $d$ of the linear code that admits $H$ as its parity-check matrix is at most $8$.
Thus, for $v > 45$, the minimum distance of $R_e\big(H'\big)$ is too high to contain the nonzero minimum weight codewords of the linear code whose parity-check matrix is $H'$.
Hence, for $v > 45$, the resulting entanglement-assisted quantum error-correcting code $\mathcal{C}$ is non-degenerate.

Now a simple counting argument shows that for $v \leq 45$ an even-replicate $S(2,3,v)$ exits only when $v = 9$, $13$, $21$, $25$, $33,$ $37$, and $45$.
If $\frac{v-1}{2} \equiv 0 \pmod{6}$, Theorem \ref{pointcodeColbourn} dictates that the minimum distance of the corresponding point code is at least $\frac{v+35}{6}$.
Hence, by the same argument as in the case when $v \geq 45$,
the two cases when $v = 25$ and $v = 37$ produce non-degenerate entanglement-assisted quantum error-correcting codes as required.
Note that by Theorem \ref{PaschSTS} an $S(2,3,13)$ is merely $3$-even-free.
Hence, the case when $v = 13$ is also settled by the same token.

The remaining case is when $v = 9$.
It is known that up to isomorphism there exists only one $S(2,3,9)$, which is the affine plane AG$(2,3)$ \cite{Beth:1999}.
Thus, by Theorem \ref{affineerasure} it is $5$-even-free but not $6$-even-free.
Hence, we only need to prove that $R_e\big(H'\big)$ does not contain a vector of weight $6$.
We first consider the sum of an even number of rows of $H'$ except the last row $(J_{1,9}, \boldsymbol{0}_{1,12})$.
Because of the $9 \times 9$ identity matrix $I$ on the left of $\left[\begin{array}{cc}I & H\end{array}\right]$, such a linear combination can be of weight $6$ or smaller
only if it is the sum of either $2$, $4$, or $6$ rows.
The sum of any pair of rows of $H$ is of weight $2\cdot4-2 = 6$ while by Theorem \ref{pointcodeColbourn}
a linear combination of rows of $H$ can be of weight less than $6$
only when it is the sum of $6$ rows, in which case it is of weight at least $2$.
Hence, no linear combination of an even number of rows results in a vector of weight $6$ in this case.
Next, we consider the sum of an even number of rows of $H'$ involving the last row $(J_{1,9}, \boldsymbol{0}_{1,12})$.
Considering the $10 \times 9$ submatrix on the left of $H'$,
the sum can be of weight $6$ or smaller only if it is the sum of either $4$, $6$, or $8$ rows,
where the contribution of rows of the submatrix to the weight of the sum is either $6$, $4$, or $2$, respectively.
By Theorem \ref{pointcodeColbourn}, the sum of $3$ or $5$ rows of $H$ is not of weight less than $4$.
Hence, the remaining case is when $7$ rows of $H'$ is added up together with $(J_{1,3^m}, \boldsymbol{0}_{1,12})$.
However, by Theorem \ref{pointcodeTonchev} a linear combination of an odd number of rows of $H$ can be of weight $4$
only when it is a single row of $H$ or the sum of $5$ rows.
Therefore, no linear combination of an even number of rows of $H'$ is of weight $6$ regardless of whether $(J_{1,3^m}, \boldsymbol{0}_{1,12})$ is involved.
This completes the proof.
\end{IEEEproof}

It is notable that from the argument in the above proof it is straightforward to see that
the entanglement-assisted quantum error-correcting code constructed
from the classical parity-check matrix $\left[\begin{array}{cc}I & H\end{array}\right]$ is also non-degenerate
when $H$ is an incidence matrix of the even-replicate Steiner $2$-design of order $v \not = 21, 33, 45$.
Hence, the operation of adding a special row done in Theorem \ref{addR} indeed improves the true distances of those quantum LDPC codes.

As we have seen throughout this paper,
assistance from reliable qubits is of coding theoretic interest,
and seems to have the potential to greatly widen the range of effectively exploitable classical error-correcting codes.
We hope that this work stimulates further studies on taking fuller advantage of classical coding theory
and also helps find interesting error correction schemes that make use of phenomena unique to the world of quantum information.

\section*{Acknowledgments}
The authors would like to thank the anonymous referees and Associate Editor Alexei Ashikhmin for their careful reading of the manuscript and valuable suggestions.

%\bibliographystyle{IEEEtran}
%\bibliography{Papers}


\end{document}